\documentclass[12pt]{article}

\usepackage{amsmath}
\usepackage{amssymb}
\usepackage{latexsym}
\usepackage{graphicx}
\usepackage{epsfig}

\usepackage{tabularx} 
\usepackage{amsmath}  
\usepackage{graphicx} 
\usepackage{cite} 
\usepackage[final]{hyperref} 
\hypersetup{
	colorlinks=true,       
	linkcolor=blue,        
	citecolor=blue,        
	filecolor=magenta,     
	urlcolor=blue         
	}

\DeclareMathOperator{\sign}{sign}

\def\beq{\begin{equation}}
\def\eeq{\end{equation}}
\def\bea{\begin{eqnarray}}
\def\eea{\end{eqnarray}}

\begin{document}

\begin{titlepage}

\vspace*{1cm}
\begin{center}
{\bf \Large Black-Hole Solutions with Scalar Hair\\[2mm] in Einstein-Scalar-Gauss-Bonnet Theories}

\bigskip \bigskip \medskip

{\bf G. Antoniou}$^{\,(a,b)}$\,\footnote{Email: anton296@umn.edu},  
{\bf A. Bakopoulos}$^{\,(b)}$\,\footnote{Email: abakop@cc.uoi.gr} and
{\bf P. Kanti}$^{\,(b)}$\,\footnote{Email: pkanti@cc.uoi.gr}

\bigskip
$^{(a)}${\it School of Physics and Astronomy, University of Minnesota,
Minneapolis,\\ MN 55455, USA}

$^{(b)}${\it Division of Theoretical Physics, Department of Physics,\\
University of Ioannina, Ioannina GR-45110, Greece}

\bigskip \medskip
{\bf Abstract}
\end{center}
In the context of the Einstein-scalar-Gauss-Bonnet theory, with a general
coupling function between the scalar field and the quadratic Gauss-Bonnet
term, we investigate the existence of regular black-hole solutions with
scalar hair. Based on a previous theoretical analysis, that studied the
evasion of the old and novel no-hair theorems, we consider a variety of
forms for the coupling function (exponential, even and odd polynomial,
inverse polynomial, and logarithmic) that, in conjunction with the profile
of the scalar field, satisfy a basic constraint. Our numerical
analysis then always leads to families of regular, asymptotically-flat
black-hole solutions with non-trivial scalar hair. The solution for the
scalar field and the profile of the corresponding energy-momentum tensor,
depending on the value of the coupling constant, may exhibit a non-monotonic
behaviour, an unusual feature that highlights the limitations of the
existing no-hair theorems. We also determine and study in detail the
scalar charge, horizon area and entropy of our solutions.

\end{titlepage}

\setcounter{page}{1}

\section{Introduction}

The construction of generalised gravitational theories, with the inclusion
of extra fields or higher-curvature terms in the action, has attracted an
enormous interest during the last decades \cite{Stelle,General}. The main reason
is that these theories may provide the framework for the ultimate theory of
gravity in the  context of which several problems of the traditional
General Relativity may be resolved. Therefore, in the context of these
modified gravitational theories, different aspects of gravity, from
black-hole solutions to  cosmological solutions, have been re-addressed
and, in several occasions, shown to lead to novel, interesting solutions. 

Such a class of solutions was the one describing regular black holes 
with a non-trivial scalar field in the exterior region, a type of solutions
forbidden by General Relativity. In an attempt to construct a criterion
whose fulfillment or violation could, respectively, forbid or allow the emergence
of such solutions, no-hair theorems were formed. In its first form, the 
{\it old no-hair theorem} \cite{NH-scalar} excluded static black holes with a
scalar field; however, this was soon outdated by the discovery of black holes
with Yang-Mills \cite{YM}, Skyrme fields \cite{Skyrmions} or with a conformal
coupling to gravity \cite{Conformal}. These led to the formulation of the
{\it novel no-hair theorem} \cite{Bekenstein} that was recently extended to
cover the case of standard scalar-tensor theories \cite{SF}; a new
form was also proposed, that covers the case of Galileon fields \cite{HN}. 

However, in a limited number of theories, even the novel no-hair theorem 
may be evaded. The first counter-example appeared very soon after its
formulation, and demonstrated the existence of regular black holes with a
scalar hair in the context of the Einstein-Dilaton-Gauss-Bonnet theory
\cite{DBH} (for some earlier studies that paved the way, see
\cite{Gibbons, Callan, Campbell, Mignemi, Kanti1995}). Additional solutions
in the context of the same theory were derived in the presence of a
Yang-Mills field \cite{Torii, KT}, in an arbitrary number of dimensions
\cite{Guo} or in the case of rotation
\cite{Kleihaus, Pani, Herdeiro, Ayzenberg} (for a number of interesting
reviews on the topic, see \cite{Charmousis, Herdeiro-review, Blazquez}).
The more recent form of the novel no-hair theorem \cite{HN} was also
shown to be evaded \cite{SZ} and concrete solutions were constructed
\cite{Babichev, Benkel}. 

A common feature of the theories that evaded the no-hair theorems was
the presence of higher-curvature terms, such as the quadratic Gauss-Bonnet
(GB) term inspired by the string theory \cite{Metsaev} or Horndeski 
theory \cite{Horndeski}. It is the presence of such terms that invalidate basic
requirements of the no-hair theorems, and open the way for the construction
of black-hole solutions with scalar hair. In a recent work of ours \cite{ABK1},
we considered a general class of Einstein-scalar-GB theories, of which the cases
\cite{DBH, SZ} constitute particular examples. We demonstrated that, under
certain constraints on the form of the coupling function between the scalar
field and the Gauss-Bonnet term, and in conjunction with the profile of
the scalar field itself, a regular black-hole horizon regime and an
asymptotically-flat regime may be smoothly connected, and thus the
no-hair theorems may be evaded. A number of novel black-hole solutions
with scalar hair was thus determined and briefly presented \cite{ABK1}.

In the present work, we provide additional support to the arguments presented
in \cite{ABK1}. We consider a general gravitational theory containing
the Ricci scalar, a scalar field and the GB term, with the latter two
quantities being coupled together through a coupling function $f(\phi)$.
Guided by the findings of our previous work \cite{ABK1}, we impose the
aforementioned constraints on the coupling function $f$ and the scalar
field $\phi$, and investigate the existence of regular, black-hole
solutions with a non-trivial scalar hair. We find a large number of
such solutions for a variety of forms for the coupling function:
exponential, polynomial (even and odd), inverse polynomial (even and
odd) and logarithmic. In all cases, the solutions for the metric
components, scalar field, curvature invariant quantities and components
of the energy-momentum tensor are derived and discussed. Further
characteristics of the produced solutions, such as the scalar charge,
horizon area and entropy, are also determined, studied in detail and
compared to the corresponding Schwarzschild values.


\section{The Theoretical Framework}

We start with the following action functional that describes a general
class of higher-curvature gravitational theories \cite{ABK1}:
\begin{equation}
S=\frac{1}{16\pi}\int{d^4x \sqrt{-g}\left[R-\frac{1}{2}\,\partial_{\mu}\phi\partial^{\mu}\phi+f(\phi)R^2_{GB}\right]}.
\label{action}
\end{equation}
In this, the Einstein-Hilbert term given by the Ricci scalar curvature $R$ is 
accompanied by the quadratic Gauss-Bonnet (GB) term $R^2_{GB}$ defined as 
\begin{equation}\label{GB-def}
R^2_{GB}=R_{\mu\nu\rho\sigma}R^{\mu\nu\rho\sigma}-4R_{\mu\nu}R^{\mu\nu}+R^2\,,
\end{equation}
in terms of the Riemann tensor $R_{\mu\nu\rho\sigma}$, Ricci tensor $R_{\mu\nu}$
and the Ricci scalar $R$. A scalar field $\phi$ also appears in the action (\ref{action}),
and couples to the GB term through a general coupling function $f(\phi)$. This
is a necessary requirement in order for the GB term, that is a total derivative in
four dimensions, to contribute to the field equations.

The gravitational field equations and the equation for the scalar field may be
derived by varying the action (\ref{action}) with respect to the metric tensor $g_{\mu\nu}$
and the scalar field $\phi$, respectively. These have the form:
\begin{equation}
G_{\mu\nu}=T_{\mu\nu}\,, \label{field-eqs}
\end{equation}
\begin{equation}
\nabla^2 \phi+\dot{f}(\phi)R^2_{GB}=0\,, \label{phi-eq_0}
\end{equation}
where $G_{\mu\nu}$ is the Einstein tensor and $T_{\mu\nu}$ the energy-momentum tensor.
The latter receives contributions from both the scalar field and the Gauss-Bonnet term
and is given by
\begin{equation}\label{Tmn}
T_{\mu\nu}=-\frac{1}{4}g_{\mu\nu}\partial_{\rho}\phi\partial^{\rho}\phi+\frac{1}{2}\partial_{\mu}\phi\partial_{\nu}\phi-\frac{1}{2}\left(g_{\rho\mu}g_{\lambda\nu}+g_{\lambda\mu}g_{\rho\nu}\right)
\eta^{\kappa\lambda\alpha\beta}\tilde{R}^{\rho\gamma}_{\quad\alpha\beta}
\nabla_{\gamma}\partial_{\kappa}f(\phi)\,,
\end{equation}
with
\begin{equation}\label{g}
\tilde{R}^{\rho\gamma}_{\quad\alpha\beta}=\eta^{\rho\gamma\sigma\tau}
R_{\sigma\tau\alpha\beta}=\frac{\epsilon^{\rho\gamma\sigma\tau}}{\sqrt{-g}}\,
R_{\sigma\tau\alpha\beta}\,.
\end{equation}
Throughout this work, the dot over the coupling function denotes its derivative
with respect to the scalar field (i.e. $\dot f =df/d\phi$), and we employ units
in which $G=c=1$.

Our aim is to find solutions of the set of Eqs. (\ref{field-eqs})-(\ref{phi-eq_0}) 
that describe regular, static, asymptotically-flat black-hole solutions with a non-trivial
scalar field. In particular, we will assume that the line-element takes the following
spherically-symmetric form
\begin{equation}\label{metric}
{ds}^2=-e^{A(r)}{dt}^2+e^{B(r)}{dr}^2+r^2({d\theta}^2+\sin^2\theta\,{d\varphi}^2)\,,
\end{equation}
and that the scalar field is also static and spherically-symmetric, $\phi=\phi(r)$.
In our quest for the aforementioned solutions, we will consider a variety of
coupling functions $f(\phi)$, that will however need to obey certain constraints
\cite{ABK1}.  

By employing the line-element (\ref{metric}), we may obtain the explicit forms of
Eqs. (\ref{field-eqs})-(\ref{phi-eq_0}): the $(tt)$, $(rr)$ and $(\theta\theta)$
components of Einstein's equations then, respectively, read
\beq
4e^B(e^{B}+rB'-1)=\phi'^2\bigl[r^2e^B+16\ddot{f}(e^B-1)\bigr]
-8\dot{f}\left[B'\phi'(e^B-3)-2\phi''(e^B-1)\right],\label{tt-eq}
\eeq
%

\vskip -0.4cm
\beq
4e^B(e^B- r A'-1)=-\phi'^2 r^2 e^B +8\left(e^B-3\right)\dot{f}A'\phi'\,,
\label{rr-eq}
\eeq
\begin{eqnarray}
&& \hspace*{-1.5cm} e^B\bigl[r{A'}^2-2B'+A'(2-rB')+2rA''\bigr]=-\phi'^2\bigl(re^B-8\ddot{f}A'\bigr)
+ \nonumber \\[1mm] 
&&\hspace*{6cm} +4\dot{f}\left[A'^2\phi'+2\phi'A''+A'(2\phi''-3B'\phi')\right],
\label{thth-eq}
\end{eqnarray}
while the equation for the scalar field takes the form:
\begin{equation}
2r\phi''+(4+rA'-rB')\,\phi'+\frac{4\dot{f}e^{-B}}{r}\left[(e^B-3)A'B'-(e^B-1)(2A''+A'^2)\right]=0\,. \label{phi-eq}
\end{equation}
In the above, the prime in the metric functions $A$ and $B$ and the scalar field $\phi$ denotes 
their differentiation with respect to the radial coordinate $r$.

Before being able to numerically integrate the system of equations (\ref{tt-eq})-(\ref{phi-eq}),
we need to reduce the number of independent variables. We observe that the $(rr)$-component
(\ref{rr-eq}) may take the form of a second-order polynomial with respect to $e^B$, i.e.
$e^{2B}+\beta e^{B}+\gamma=0$. Therefore, this can be solved to give:
\begin{equation}\label{Bfunction}
e^B=\frac{-\beta\pm\sqrt{\beta^2-4\gamma}}{2}\,,
\end{equation}
where
\beq
\beta=\frac{r^2{\phi'}^2}{4}-(2\dot{f}\phi'+r) A'-1\,, \qquad 
\gamma=6\dot{f}\phi'A'\label{bg}\,.
\eeq
According to the above, the solution for the metric function $B(r)$ may be easily
found once the solutions for the scalar field $\phi(r)$ and the metric function $A(r)$
are determined. In the remaining system of equations (\ref{tt-eq}), (\ref{thth-eq})
and (\ref{phi-eq}), $e^B$ may therefore be eliminated by using Eq. (\ref{Bfunction})
while $B'$ may be replaced by the expression
\begin{equation}\label{B'}
B'=-\frac{\gamma'+\beta' e^{B}}{2e^{2B}+\beta e^{B}}\,,
\end{equation}
that follows by differentiating Eq. (\ref{Bfunction}) with respect to the radial coordinate. 
Subsequently, the remaining three equations (\ref{tt-eq}), (\ref{thth-eq}) and
(\ref{phi-eq}) form a
system of only two independent, ordinary differential equations of second order for
the functions $A$ and $\phi$:
\begin{align} \label{A}
A''=&\frac{P}{S}\,,\\
\phi''=&\frac{Q}{S}\,. \label{phi}
\end{align} 
In the above equations, $P$, $Q$ and $S$ are complicated expressions of
$(r, e^B, \phi', A', \dot f, \ddot f)$ that are given in the Appendix. Note, that in these
expressions we have eliminated, via Eq. (\ref{B'}), $B'$ that involves
$A''$ and $\phi''$, but retained $e^B$ for notational simplicity.

\subsection{Asymptotic Solution at Black-Hole Horizon}

We will start our quest for black-hole solutions with a non trivial scalar hair by
determining first the asymptotic solutions of the set of Eqs. (\ref{tt-eq})-(\ref{phi-eq})
near the black-hole horizon and at asymptotic infinity. These solutions will serve as
boundary conditions for our numerical integration but will also provide important
constraints on our theory (\ref{action}). Near the black-hole horizon $r_h$, it is
usually assumed that the metric functions and the scalar field may be expanded as
\begin{align}
e^{A}&=\sum_{n=1}^\infty{a_n(r-r_h)^n}\,,\label{expA_rh}\\
e^{-B}&=\sum_{n=1}^\infty{b_n(r-r_h)^n}\,,\label{expB_rh}\\
\phi &=\sum_{n=0}^{\infty}\frac{\phi^{(n)}(r_h)}{n!}\,(r-r_h)^n\,, \label{phi_rh}
\end{align}
where $(a_n, b_n)$ are constant coefficients and $\phi^{(n)}(r_h)$ denotes the ($n$th)-derivative
of the scalar field evaluated at the black-hole horizon. Equations (\ref{expA_rh})-(\ref{expB_rh})
reflect the expected behaviour of the metric tensor near the horizon of a spherically-symmetric
black hole with the solution being regular if the scalar coefficients $\phi^{(n)}(r_h)$ in
Eq. (\ref{phi_rh}) remain finite at the same regime. 

In a recent work of ours \cite{ABK1}, we have followed instead the alternative approach 
of assuming merely that, near the horizon, the metric function $A(r)$ diverges, in 
accordance to Eq. (\ref{expA_rh}). Then, the system of differential equations was evaluated
in the limit $r \rightarrow r_h$, and the finiteness of the quantity $\phi''$ was demanded.
This approach was followed in \cite{DBH} where an exponential coupling function was
assumed between the scalar field and the GB term. In \cite{ABK1}, the form of the
coupling function $f(\phi)$ was left arbitrary, and the requirement of the finiteness of
$\phi''_h$ was shown to be satisfied only under the constraint

\begin{equation}\label{constraint1}
r_h^3\phi'_h+12\dot{f}_h+2r_h^2{\phi'}_h^2\dot{f}_h=0\,,
\end{equation}
where all quantities have been evaluated at the horizon $r_h$. The above is a 
second-order polynomial with respect to $\phi'_h$, and may be easily solved to yield
the solutions:
\begin{equation}\label{con-phi'}
\phi'_h=\frac{r_h}{4\dot{f}_h}\left(-1\pm\sqrt{1-\frac{96\dot{f}_h^2}{r_h^4}}\right).
\end{equation}
The above ensures that an asymptotic black-hole solution with a regular scalar field 
exists for a general class of theories of the form (\ref{action}). The only constraint
on the form of the coupling function arises from the demand that the first derivative
of the scalar field on the horizon must be real, which translates to the inequality
\begin{equation}\label{con-f}
\dot{f}_h^2<\frac{r_h^4}{96}\,.
\end{equation}
Assuming the validity of the constraint (\ref{constraint1}), Eq. (\ref{A}) then uniquely
determines the form of the metric function $A$ in the near-horizon regime; through
Eq. (\ref{Bfunction}), the metric function $B$ is also determined. Therefore, the
asymptotic solution of the Eqs. (\ref{Bfunction}), (\ref{A}) and (\ref{phi}), in the
limit $r \rightarrow r_h$, is given by the expressions
\bea
&&e^{A}=a_1 (r-r_h) + ... \,, \label{A-rh}\\[1mm]
&&e^{-B}=b_1 (r-r_h) + ... \,, \label{B-rh}\\[1mm]
&&\phi =\phi_h + \phi_h'(r-r_h)+ \phi_h'' (r-r_h)^2+ ... \,. \label{phi-rh}
\eea
and describes, by construction, a black-hole horizon with a regular scalar field
provided that $\phi'$ obeys the constraint (\ref{con-phi'}) and the coupling
function $f$ satisfies Eq. (\ref{con-f}). We note that the desired form of the
asymptotic solution was derived only for the choice of the $(+)$-sign in
 Eq. (\ref{Bfunction}) as the $(-)$-sign fails to lead to a black-hole solution \cite{ABK1}.

The aforementioned regularity of the near-horizon solution should be reflected 
to the components of the energy-momentum tensor $T_{\mu\nu}$ as well as to the scalar 
invariant quantities of the theory. The non-vanishing components of the energy momentum
tensor (\ref{Tmn}) are:
\begin{align}
T^t_{\;\,t}=&-\frac{e^{-2B}}{4r^2}\left[\phi'^2\left(r^2e^B+16\ddot{f}(e^B-1)\right)-8\dot{f}\left(B'\phi'(e^B-3)-2\phi''(e^B-1)\right)\right],\label{Ttt}\\
T^r_{\;\,r}=&\frac{e^{-B}\phi'}{4}\left[\phi'-
\frac{8e^{-B}\left(e^B-3\right)\dot{f}A'}{r^2}\right],\label{Trr}\\
T^{\theta}_{\;\,\theta}=&T^{\varphi}_{\;\,\varphi}=-\frac{e^{-2B}}{4 r}\left[\phi'^2\left(re^B-8\ddot{f}A'\right)-4\dot{f}\left(A'^2\phi'+2\phi'A''+
A'(2\phi''-3B'\phi')\right)\right].\label{Tthth}
\end{align}
Employing the asymptotic behaviour given in Eqs. (\ref{A-rh})-(\ref{phi-rh}), we readily
derive the following approximate behaviour:
\begin{align}
T^t_{\;\,t}=&+\frac{2e^{-B}}{r^2}B'\phi'\dot{f}+\mathcal{O}(r-r_h)\,,\label{Ttt-rh}\\
T^r_{\;\,r}=&-\frac{2e^{-B}}{r^2}A'\phi'\dot{f}+\mathcal{O}(r-r_h)\,,\label{Trr-rh}\\
T^{\theta}_{\;\,\theta}=&T^{\varphi}_{\;\,\varphi}=\frac{e^{-2B}}{r}\phi' \dot{f}\left(2A''+A'^2-3A'B'\right)
+\mathcal{O}(r-r_h)\,.\label{Tthth-rh}
\end{align}
The above expressions, in the limit $r \rightarrow r_h$, lead to constant values for all
components of the energy-momentum tensor. Similarly, one may see that the scalar
invariant quantities
$R$, $R_{\mu\nu}R^{\mu\nu}$ and $R_{\mu\nu\rho\sigma} R^{\mu\nu\rho\sigma}$,
whose exact expressions are listed in the Appendix, reduce to the approximate forms
\begin{align}
R = &+\frac{2 e^{-B}}{r^2} \left(e^{B}-2 r A'\right) +\mathcal{O}(r-r_h)\,,\\
R_{\mu\nu}R^{\mu\nu}= &+\frac{2e^{-2 B}}{r^4} \left(e^{B} -r A'\right)^2+\mathcal{O}(r-r_h)\,,\\
R_{\mu\nu\rho\sigma}R^{\mu\nu\rho\sigma} =&+\frac{4 e^{-2 B}}{r^4} \left( e^{2B} + 
r^2 A'^2\right)+\mathcal{O}(r-r_h)\,.
\end{align}
In the above, we have used that, near the horizon, $A' \approx -B' \approx 1/(r-r_h)$
and $A'^2 \approx -A''$,
as dictated by Eqs. (\ref{A-rh})-(\ref{B-rh}). Again, the dominant term in each curvature
invariant adopts a constant, finite value in the limit $r \rightarrow r_h$. Subsequently,
the GB term also comes out to be finite, in the same limit, and given by
\beq
R^2_{GB} = +\frac{12  e^{-2 B}}{r^4}A'^2 +\mathcal{O}(r-r_h)\,. \label{GB-rh}
\eeq


\subsection{Asymptotic Solution at Infinity}

At the other asymptotic regime, that of radial infinity ($r \rightarrow \infty$), the 
metric functions and the scalar field may be again expanded in power series,
this time in terms of $1/r$. Demanding that the metric components reduce to
those of the asymptotically-flat Minkowski space-time while the scalar field assumes
a constant value, we write:
\begin{align}
e^{A}&=1+\sum_{n=1}^\infty{\frac{p_n}{r}}\,,\label{Afar}\\
e^B&=1+\sum_{n=1}^\infty{\frac{q_n}{r}}\,,\label{Bfar}\\
\phi &=\phi_{\infty}+\sum_{n=1}^{\infty}{\frac{d_n}{r}}\,.\label{phifar}
\end{align}
The arbitrary coefficients $(p_n, q_n, d_n)$ are in principle determined upon substitution
of Eqs. (\ref{Afar})-(\ref{phifar}) in the field equations of the theory. However, two of
the coefficients, namely $p_1$ and $d_1$, remain as free parameters and are associated
with the Arnowitt-Deser-Misner (ADM) mass and scalar charge, respectively: 
$p_1 \equiv -2M$ and $d_1=D$.
The remaining coefficients may be calculated to an arbitrary order: we have performed
this calculation up to order $\mathcal{O}(1/r^6)$, and derived the following expressions:
\begin{align}
e^A=&\; 1-\frac{2M}{r}+\frac{MD^2}{12r^3}+\frac{24MD\dot{f}+M^2D^2}{6r^4}\nonumber\\
&-\frac{96M^3D-3MD^3+512M^2\dot{f}-64D^2\dot{f}+128MD^2\ddot{f}}{90r^5}+\mathcal{O}(1/r^6)\,,\\
e^B=&\; 1+\frac{2M}{r}+\frac{16M^2-D^2}{4r^2}+\frac{32M^3-5MD^2}{4r^3}+\frac{768M^4-208M^2D^2-384MD\dot{f}+3D^4}{48r^4}\nonumber\\
&+\frac{6144M^5-2464M^3D^2+97MD^4-6144M^2D\dot{f}+192D^3\dot{f}-384MD^2\ddot{f}}{192r^5}\nonumber\\
&+\mathcal{O}(1/r^6)\,,\\
\phi=&\; \phi_{\infty}+\frac{D}{r}+\frac{MD}{r^2}+\frac{32M^2D-D^3}{24r^3}+\frac{12M^3D-24M^2\dot{f}-MD^3}{6r^4}\nonumber\\
&+\frac{6144M^4D-928M^2D^3+9D^5
-12288M^3\dot{f}-1536MD^2\dot{f}}{1920r^5}+\mathcal{O}(1/r^6)\,.
\end{align}
We observe that the scalar charge $D$ modifies significantly the expansion
of the metric functions at order $\mathcal{O}(1/r^2)$ and higher. The
existence itself of $D$, and thus of a non-trivial form for the scalar field,
is caused by the presence of the GB term in the theory. The exact form,
however, of the coupling function does not enter in the above expansions
earlier than the order $\mathcal{O}(1/r^4)$. This shows that an asymptotically
flat solution of Eqs. (8)-(11), with a constant scalar field does not require a
specific coupling function and, in fact, arises for an arbitrary form of this function. 

The asymptotic solution at infinity, given by Eqs. (\ref{Afar})-(\ref{phifar}), is also
characterised by regular components of $T_{\mu\nu}$ and curvature invariants.
Employing the facts that, as $r \rightarrow \infty$, $(e^A,\,e^B,\,\phi) \approx \mathcal{O}(1)$
while $(A',\,B',\,\phi') \approx \mathcal{O}\left(1/r^2\right)$, we find
for the components of the energy-momentum tensor the asymptotic behaviour:
\beq
T^t_{\;\,t} \simeq -T^r_{\;\,r} \simeq 
T^\theta_{\;\,\theta} \simeq -\frac{1}{4}\,\phi'^2 + {\mathcal O}\left(\frac{1}{r^6}\right).
\label{Tmn-far}
\eeq
Clearly, all of the above components go to zero, as expected. A similar behaviour is
exhibited by all curvature invariants and the GB term, in accordance to the 
asymptotically-flat limit derived above. In particular, for the GB term, we obtain
\beq
R^2_{GB} \approx \frac{48 M^2}{r^6}\,. \label{GB-far}
\eeq


\subsection{Connecting the two asymptotic solutions}

In the previous two subsections, we have constructed a near-horizon solution
with a regular scalar field and an asymptotically-flat solution with a constant
scalar field -- that was achieved under mild constraints on the form of
the coupling function $f(\phi)$. However, given the complexity of the
equations of the theory, it is the numerical integration of the system
(\ref{A})-(\ref{phi}) that will reveal whether these two asymptotic solutions
may be smoothly matched to create a black-hole solution with scalar hair
valid over the entire radial domain.

In fact, theoretical arguments developed decades ago, the so-called no-hair
theorems, excluded in the past the emergence of such solutions in a variety
of scalar-tensor theories. The older version of the no-hair theorem \cite{NH-scalar}
was applied in theories with minimally-coupled scalar fields: it employs
the scalar equation of motion and relies on the sign of $V'(\phi)$, where
$V(\phi)$ is the potential of the scalar field. In most theories studied,
the quantity $V'(\phi)$ had the opposite sign from the demanded one, and this
excluded the  emergence of the desired black-hole solutions. In \cite{ABK1},
the same argument was applied in the case of the theory (\ref{action}),
appropriately altered to yield a constraint on the effective potential
$V_{eff}(\phi)=f(\phi) R^2_{GB}$ of the scalar field. Although this constraint
was of an integral form over the entire radial regime, in special cases
it merely demanded that $f(\phi) R^2_{GB}>0$. In Eqs. (\ref{GB-rh}) and
(\ref{GB-far}), the asymptotic values of the GB term near the horizon
and at infinity were derived: they are both positive, with the latter
decreasing as $r$ increases. These results point to a monotonic
decreasing behaviour of the GB term from an initial positive value
near the horizon to a vanishing value at radial infinity - as we will
shortly demonstrate in the next section, this is indeed the behaviour
of the GB term. In that case, the only requirement for the evasion of 
the old no-hair theorem is apparently the positivity of the coupling function
$f(\phi)$.

As the old no-hair theorem imposes in general mild constraints on a theory,
in \cite{ABK1} we considered in addition the novel no-hair theorem \cite{Bekenstein}
that applies also in theories with a conformal coupling of the scalar fields
to gravity (see, also, \cite{SF}). This argument relies on the profile of the
$T^r_{\;\,r}$ component of the energy-momentum tensor of the theory in terms
of the radial coordinate. In a large class of theories, under the assumptions
of positivity and conservation of energy,  it is in general extremely difficult
to smoothly match the near-horizon and asymptotic-infinity values of
$T^r_{\;\,r}$. That prevents the emergence of black-hole solutions
with scalar hair. However, the novel no-hair theorem was shown to
be evaded in the context of the Einstein-Scalar-Gauss-Bonnet theory
with an exponential coupling function \cite{DBH} or a linear coupling
function \cite{SZ}. In both theories, the presence of the GB term played
a catalytic role for the emergence of the solutions. Therefore, in
\cite{ABK1}, we kept again the form of the coupling function $f(\phi)$
arbitrary, and reconsidered the argument of the novel no-hair theorem. 
Our analysis revealed that the evasion of this theorem holds for a 
general class of theories involving the GB term: the profile of 
$T^r_{\;\,r}$ in terms of $r$ may be easily made smooth and monotonic
under the assumptions that, near the horizon,
\beq
\dot f \phi' <0, \qquad \dot f \phi'' + \ddot f \phi'^2>0\,.
\label{const-novel}
\eeq
The first constraint, according to Eq. (\ref{Trr-rh}) ensures the 
positivity of $T^r_{\;\,r}$ in the near-horizon regime; the second
that $(T^r_{\;\,r})'$ is negative in the same regime. Then, in
conjunction with the behaviour described by Eq. (\ref{Tmn-far})
at radial infinity, $T^r_{\;\,r}$ is positive and decreasing over
the whole radial regime. This behaviour invalidates the requirements
set by the novel no-hair theorem and thus causes its evasion. 

In fact, the first of the constraints listed in Eq. (\ref{const-novel})
is already satisfied: Eq. (\ref{con-phi'}) dictates that the combination
$\dot f \phi'$ at the horizon is always negative to ensure the regularity
of the black-hole horizon. Therefore, if $\dot f >0$, then $\phi'_h$ must
be necessarily negative, or vice versa. The remaining constraint
$\dot f \phi'' + \ddot f \phi'^2>0$ may be alternatively written as
$\partial_r (\dot f \phi')|_{r_h}>0$; this merely demands that the
aforementioned negative value of the quantity $(\dot f \phi')|_{r_h}$ should be 
constrained away from the horizon so that the two asymptotic solutions
(\ref{A-rh})-(\ref{phi-rh}) and (\ref{Afar})-(\ref{phifar}) can smoothly
match. As we will see in the next section, this second constraint is
automatically satisfied for all the solutions found and does not demand any
fine-tuning of our parameters.


\section{Numerical Solutions}

The derivation of exact solutions, valid over the entire radial domain, demands
the numerical integration of the system (\ref{A})-(\ref{phi}). Our integration
starts at a distance very close to the horizon of the black hole, i.e. at
$r\approx r_h+\mathcal{O}(10^{-5})$ (for simplicity, we set $r_h=1$). There,
we use as boundary conditions the asymptotic solution (\ref{A-rh})-(\ref{phi-rh})
together with Eq. (\ref{con-phi'}) for $\phi'_h$ upon choosing a particular coupling
function $f(\phi)$. The integration proceeds towards large values of the radial
coordinate until the form of the derived solution matches the asymptotic solution
(\ref{Afar})-(\ref{phifar}). In the next sub-sections, we present a variety of
regular black-hole solutions with scalar hair for different choices of the coupling
function $f(\phi)$. 

\begin{figure}[b!]
\begin{center}
\minipage{0.51\textwidth}
  \includegraphics[width=\linewidth]{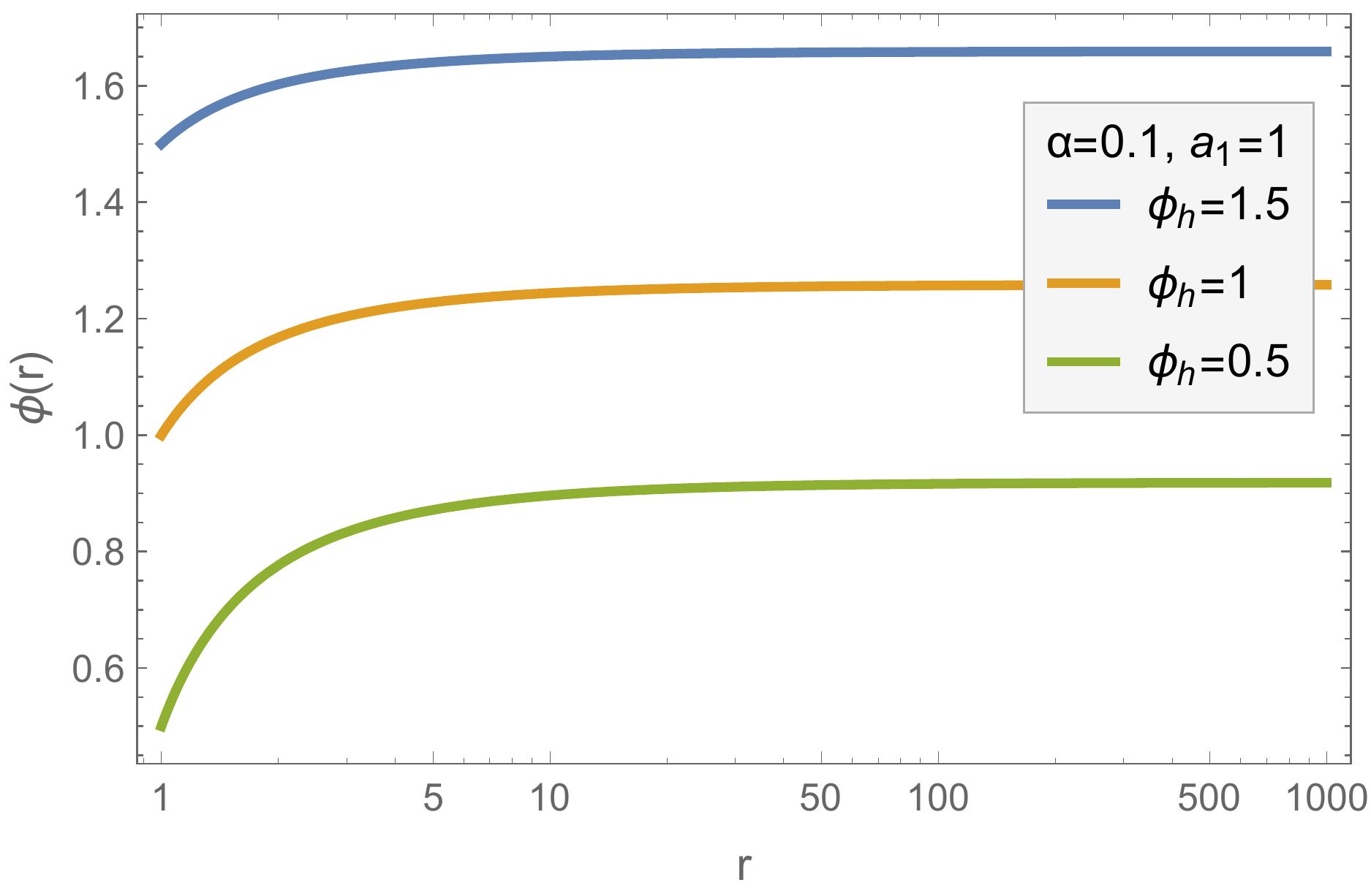}
\endminipage\hfill \hspace*{-1.5cm}
\minipage{0.51\textwidth}
  \includegraphics[width=\linewidth]{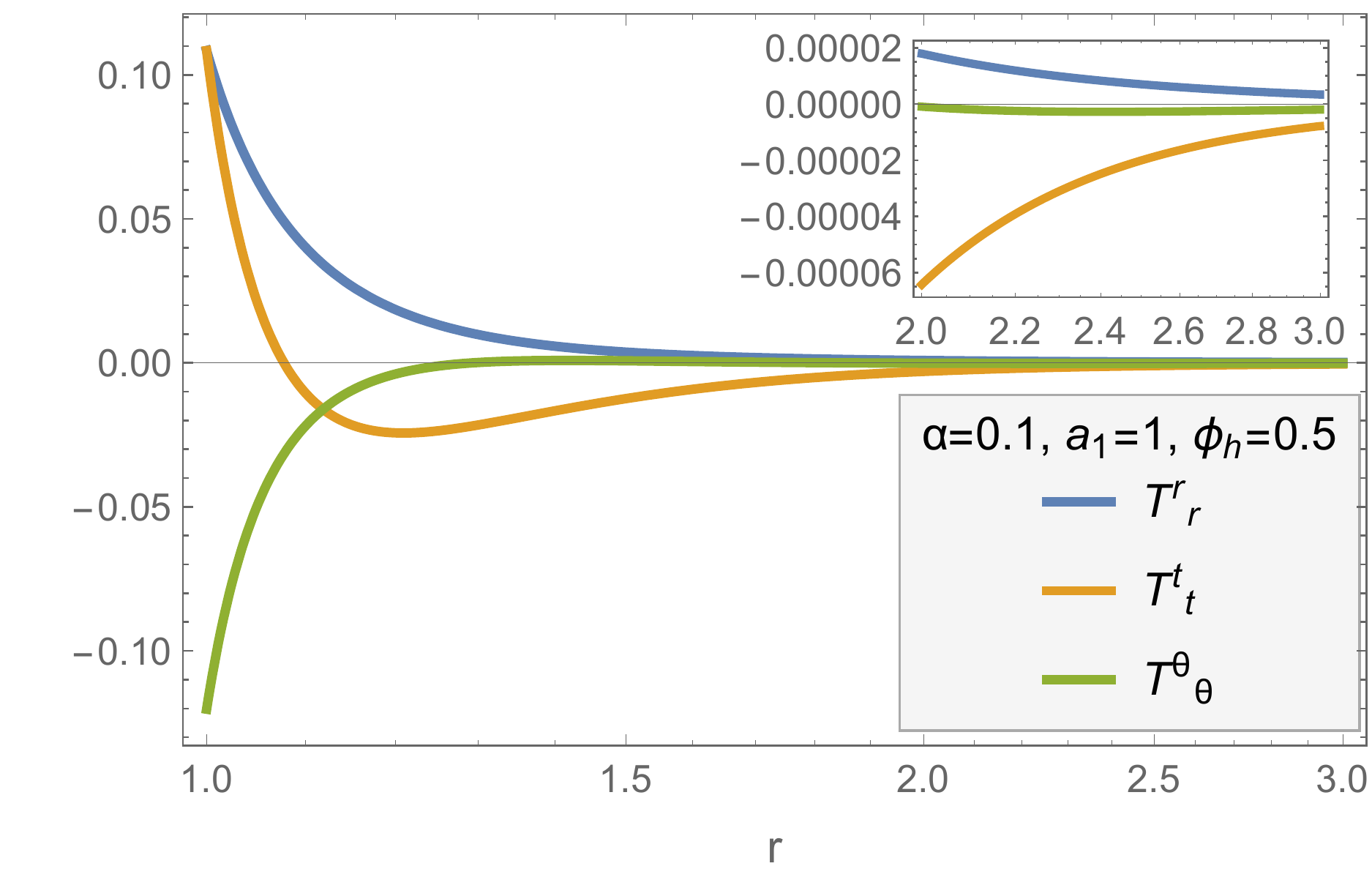}
\endminipage\hfill
    \caption{The scalar field $\phi$ (left plot), and the energy-momentum tensor
$T_{\mu\nu}$ (right plot) in terms of the radial coordinate $r$, for $f(\phi)=\alpha e^{-\phi}$.}
   \label{Phi_Exp}
  \end{center}
\end{figure}

\subsection{Exponential Coupling Function}

First, we consider the case where $f(\phi)=\alpha e^{\kappa \phi}$. According to
the arguments presented in Section 2.3, the evasion of the  old no-hair theorem
is ensured for $f(\phi)>0$; therefore, we focus on the case with $\alpha>0$. As
the exponential function is always positive-definite, the sign of 
$\dot f= \alpha \kappa e^{\kappa \phi}$ is then determined by the sign of $\kappa$.
In order to evade also the novel no-hair theorem and allow for regular black-hole
solutions to emerge, we should satisfy the constraint $\dot f \phi'<0$,
or equivalently $\kappa\,\phi'<0$, near the horizon. Therefore, for $\kappa>0$,
we should have $\phi'_h<0$, which causes the decreasing of the scalar field as
we move away from the black-hole horizon. The situation is reversed for
$\kappa<0$ when $\phi_h'>0$ and the scalar field increases with $r$. 

The case of $f(\phi)=\alpha e^\phi$, with $\alpha>0$, was studied in \cite{DBH}
and led to the well-known family of Dilatonic Black Holes. The solutions were
indeed regular and asymptotically-flat with the scalar field decreasing away
from the horizon, in agreement with the above discussion. Here, we present the
complimentary case with $f(\phi)=\alpha e^{-\phi}$ (the exact value
of $\kappa$ does not alter the physical characteristics of the solution and,
here, we set it to $\kappa=-1$). 
At the left plot of Fig. \ref{Phi_Exp}, we present a family of solutions for
the scalar field $\phi$ for different initial values $\phi_h$: for $\kappa=-1<0$,
the scalar field must necessarily have $\phi_h'>0$, and therefore increases as
$r$ increases. The value of the coupling constant $\alpha$, once the form of
the coupling function and the asymptotic value $\phi_h$ are chosen, is
restricted by the inequality (\ref{con-f}) - here,
we present solutions for indicative allowed values of $\alpha$.

\begin{figure}[t]
\begin{center}
\mbox{\hspace*{-0.9cm} \includegraphics[width = 0.54 \textwidth] {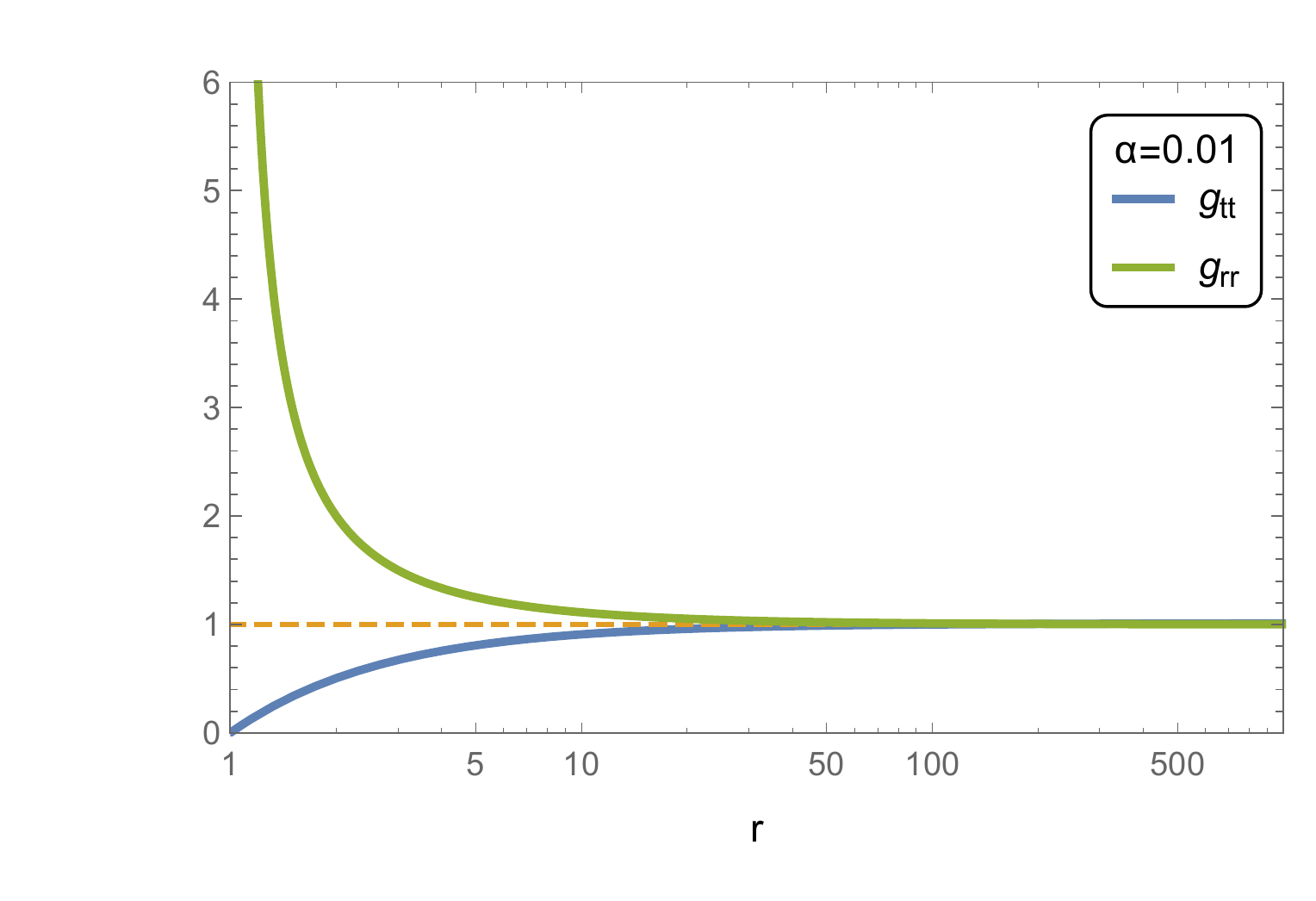}}
\hspace*{-0.9cm} {\includegraphics[width = 0.54 \textwidth]
{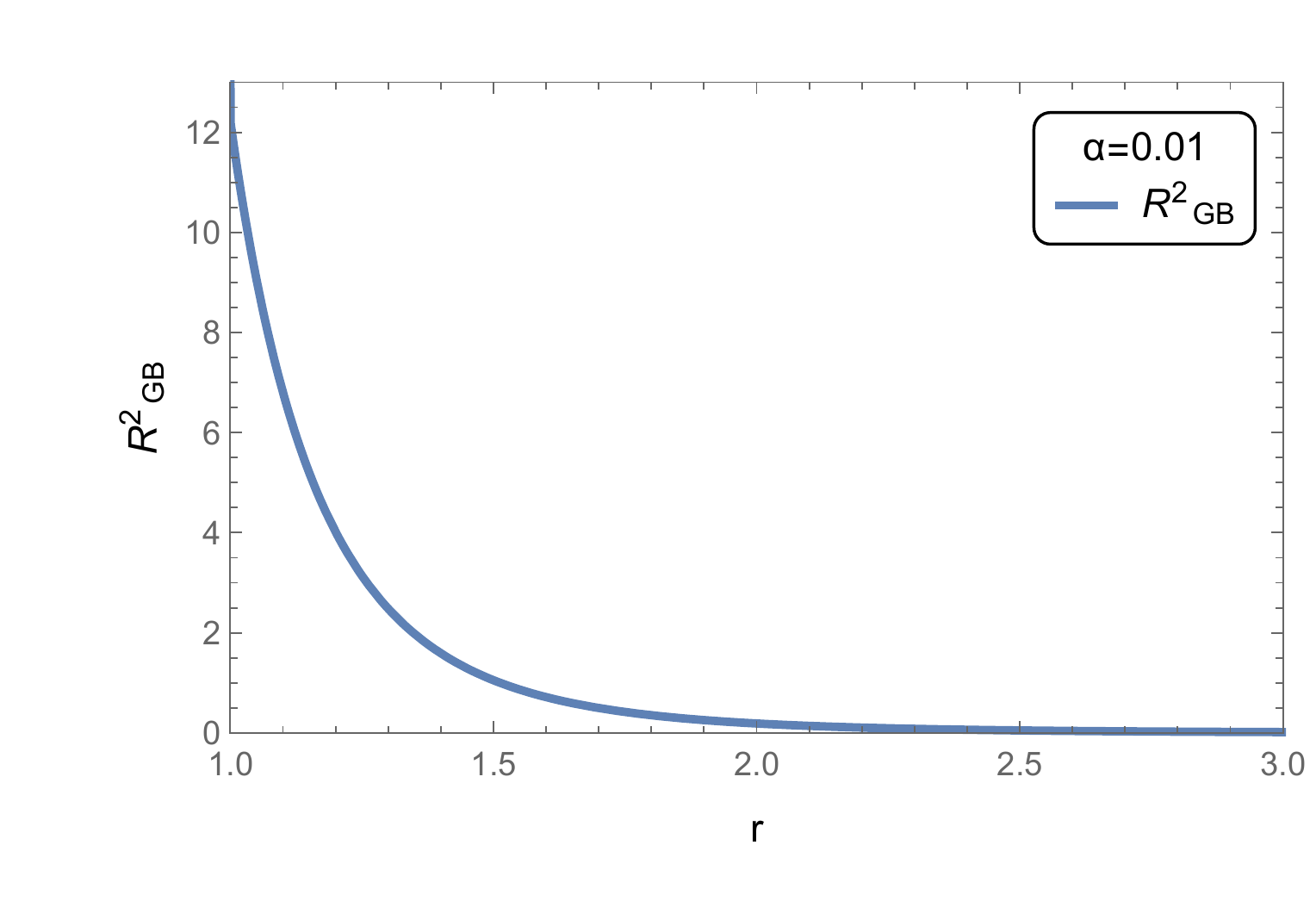}}
    \caption{The metric components $g_{tt}$ and $g_{rr}$ (left plot), and the Gauss-Bonnet
term $R^2_{GB}$ (right plot) in terms of the radial coordinate $r$, for $f(\phi)=\alpha e^{-\phi}$.}
   \label{Metric_GB}
  \end{center}
\end{figure}

At the right plot of Fig. \ref{Phi_Exp}, we present
the energy-momentum tensor components for an indicative solution of this family:
clearly, all components remain finite over the whole radial regime. In particular,
$T^r_{\,\,r}$ remains positive and monotonically decreases towards infinity,
exactly the behaviour that ensures the evasion of the novel no-hair theorem.
As discussed in the previous section, apart from choosing the input value $\phi'_h$
for our numerical integration in accordance with Eq. (\ref{con-phi'}), no
other fine-tuning is necessary; the second constraint for the evasion of the novel
no-hair theorem, i.e. $\dot f \phi'' + \ddot f \phi'^2>0$, is automatically satisfied
without any further action, and this is reflected in the decreasing behaviour of $T^r_{\,\,r}$
component near the horizon.

At the left and right plots of Fig. \ref{Metric_GB}, we also present the solution
for the two metric components ($|g_{tt}|,g_{rr}$) and the GB term $R^2_{GB}$, respectively. 
The metric components exhibit the expected behaviour near the black-hole horizon
with $g_{tt}$ vanishing and $g_{rr}$ diverging at $r_h=1$. In order to ensure
asymptotic flatness at radial infinity, the free parameter $a_1$ appearing in the
near-horizon solution (\ref{A-rh}) is appropriately chosen. On the other hand,
the GB term remains finite and
positive-definite over the entire radial domain - in fact it displays the monotonic
behaviour, hinted by its two asymptotic limits (\ref{GB-rh}) and (\ref{GB-far}),
that causes the evasion of the old no-hair theorem. As expected, it contributes
significantly near the horizon, where the curvature is large, and quickly fades
away as we move towards larger distances. The profile of the metric
components and GB term exhibit the same qualitative behaviour in all families
of black-hole solutions presented in this work, so we refrain from giving
additional plots of these two quantities in the next sub-sections. 

\begin{figure}[t]
\begin{center}
\minipage{0.49\textwidth}
  \includegraphics[width=\linewidth]{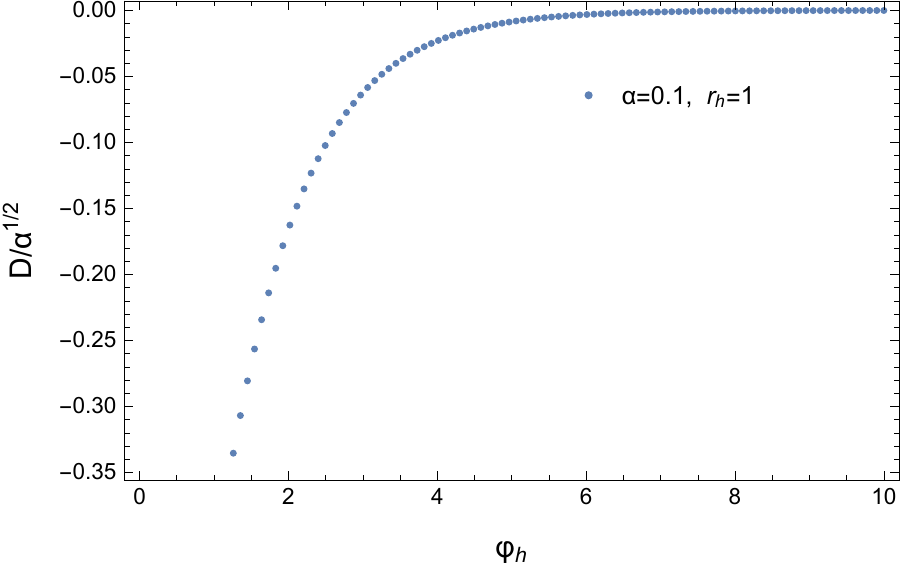}
\endminipage\hfill \hspace*{-0.4cm}
\minipage{0.49\textwidth}
  \includegraphics[width=\linewidth]{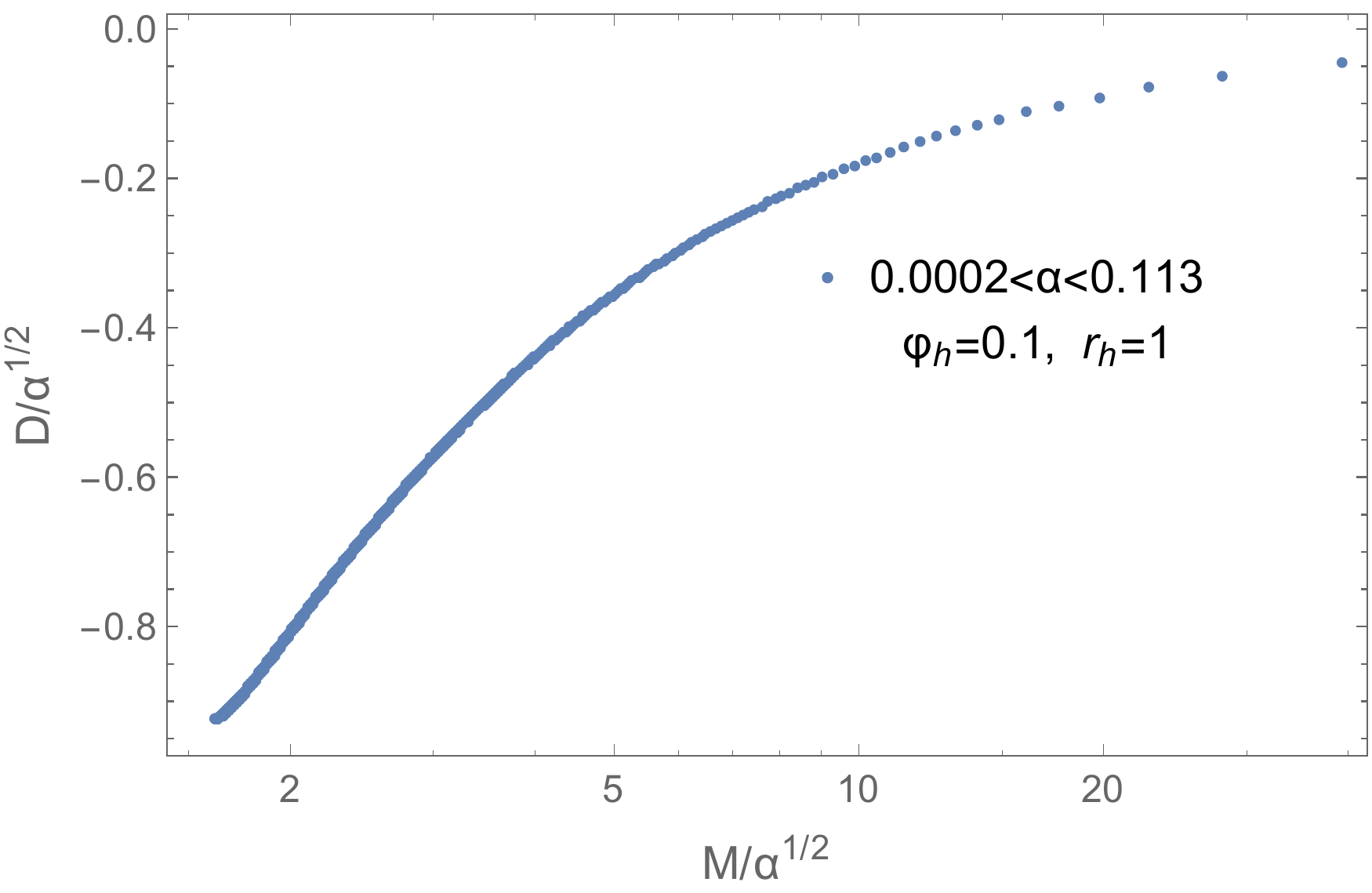}
\endminipage\hfill
    \caption{The scalar charge $D$ as a function of the near-horizon value $\phi_h$
(left plot), and of its mass $M$ (right plot), for $f(\phi)=\alpha e^{-\phi}$.}
   \label{scalarD-exp}
  \end{center}
\end{figure}

The profile of the scalar charge $D$ as a function of the near-horizon value $\phi_h$
and of the mass $M$ is given at the left and right plot, respectively, of Fig.
\ref{scalarD-exp} (each dot in these, and subsequent, plots stand for a different
black-hole solution). 
For the exponential coupling function $f(\phi)=\alpha e^{-\phi}$, and for fixed $\alpha$
and $r_h$, the scalar field near the horizon $\phi_h$ may range from a
minimum value, dictated by Eq. (\ref{con-f}), up to infinity. As the left plot reveals,
as $\phi_h \rightarrow \infty$, the coupling of the scalar field to the GB term vanishes,
and we recover the Schwarzschild case with a trivial scalar field and a vanishing charge.
On the other hand, in the right plot, we observe that as the mass of the black-hole
increases, the scalar charge decreases in absolute value, and thus larger black holes
tend to have smaller charges. 

It is also interesting to study the profiles of the area of the black-hole horizon, 
$A_h=4 \pi r_h^2$, and of the entropy $S_h$ of this class of solutions. 
The entropy is defined through the relation \cite{GH}
\beq
S_h=\beta\left[\frac{\partial (\beta F)}{\partial \beta} -F\right],
\label{entropy-def}
\eeq
where $F=I_E/\beta$ is the Helmholtz free-energy of the system given in terms of the
Euclidean version of the action $I_E$. Also, $\beta=1/(k_B T)$ with the temperature 
following easily from the definition \cite{York, GK}
\beq 
T=\frac{k}{2\pi}=\frac{1}{4\pi}\,\left(\frac{1}{\sqrt{|g_{tt} g_{rr}|}}\,
\left|\frac{dg_{tt}}{dr}\right|\right)_{r_h}=\frac{\sqrt{a_1 b_1}}{4\pi}\,.
\label{Temp-def}
\eeq
The calculation of the temperature and entropy of the dilatonic black hole, with
an exponential coupling function of the form $f(\phi)=\alpha e^\phi$, was performed
in detail in \cite{KT}. By closely repeating the analysis, we find the following expressions
for the temperature
\beq 
T=\frac{1}{4\pi}\,\frac{(2M+D)}{r_h^2+4 f(\phi_h)}\,,
\label{Temp-gen}
\eeq
and entropy 
\beq
S_h=\frac{A_h}{4} +4 \pi f(\phi_h)
\label{entropy}
\eeq
of a GB black-hole arising in the context of our theory (\ref{action}) with a general
coupling function $f(\phi)$ between the scalar field and the GB term. We easily confirm
that, in the absence of the coupling function, the above quantities reduce to the
corresponding Schwarzschild ones, $T=1/(4\pi r_h)$ and $S_h=A_h/4$, respectively.

\begin{figure}[t]
\begin{center}
\minipage{0.52\textwidth} \hspace*{-0.5cm}
  \includegraphics[width=\linewidth]{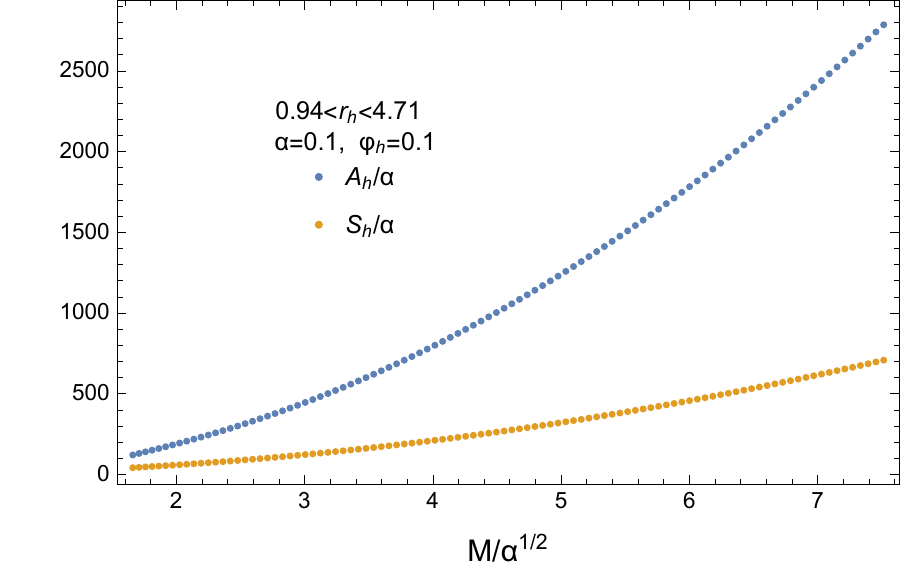}
\endminipage\hfill \hspace*{-1.5cm}
\minipage{0.52\textwidth}
  \includegraphics[width=\linewidth]{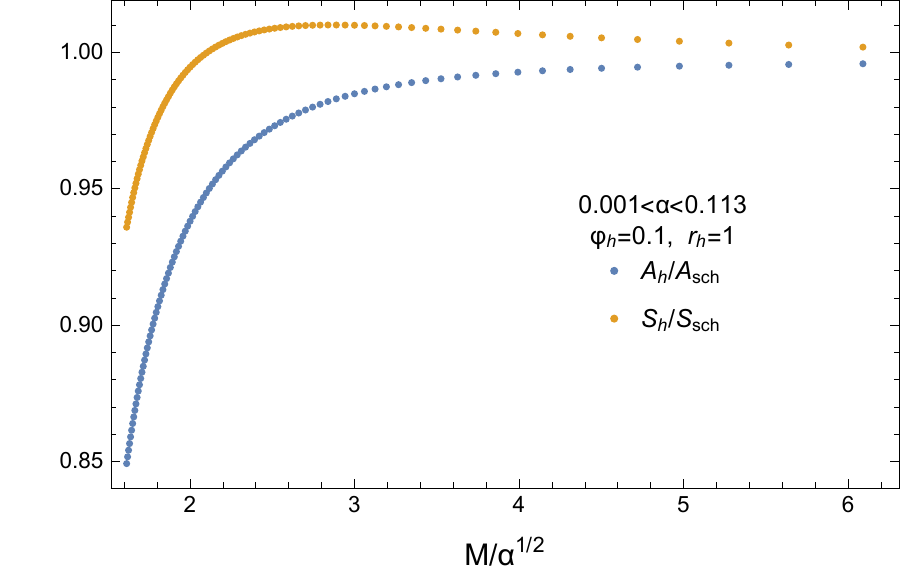}
\endminipage\hfill
    \caption{The horizon area $A_h$ and entropy $S_h$ of the black hole (left plot)
and their ratios to the corresponding Schwarzschild values (right plot) in terms of the
mass $M$ for $f(\phi)=\alpha e^{-\phi}$.}
   \label{AS-exp}
  \end{center}
\end{figure}

By employing the expressions for $A_h$ and $S_h$ as given above, we depict the horizon
area and entropy, in terms of the mass of the black hole, at the left plot of Fig. \ref{AS-exp}:
we observe that both quantities increase fast as the mass increases.
The right plot allows us to compare more effectively our solutions to the Schwarzschild
one. The lower curve depicts the ratio of $A_h$ to the area of the Schwarzschild
solution $A_{Sch}=16 \pi M^2$, as a function of $M$; we observe that, for large black-hole
masses, the ratio $A_h/A_{Sch}$ approaches unity, therefore, large GB black holes are
not expected to deviate in their characteristics from the Schwarzschild solution of the
same mass. On the other hand, in the small-mass limit, the ratio $A_h/A_{Sch}$
significantly deviates from unity; in addition, a lower bound appears for the black-hole
radius, and thus of the mass of the
black hole, due to the constraint (\ref{con-f}) not present in the Schwarzschild case --
this feature has been noted before in the case of the dilatonic black holes \cite{DBH,
Blazquez}. It is worth noting that the GB term, as an extra gravitational effect, causes
the shrinking of the size of the black hole as the ratio $A_h/A_{Sch}$ remains for
all solutions below unity. 

Turning finally to the entropy $S_h$ of our black-hole solutions, we find a similar
pattern: very small GB black holes differ themselves from the Schwarzschild solution by
having a lower entropy whereas large GB black holes tend to acquire, among other
characteristics, and the entropy of the Schwarzschild solution. We observe however
that, apart from the very small-mass regime close to the minimum value, the 
`exponential' GB
black holes have in general a higher entropy than the Schwarzschild solution, a 
characteristic that points to the thermodynamical stability of these solutions. In
fact, the dilatonic GB black holes \cite{DBH}, which comprise a subclass of this
family of solutions with a coupling function of the form $f(\phi)=\alpha e^\phi$,
have an identical entropy pattern, and were shown to be linearly stable under small
perturbations more than twenty years ago.


\subsection{Even Polynomial Function}

\begin{figure}[b!]
\minipage{0.49\textwidth}
  \includegraphics[width=\linewidth]{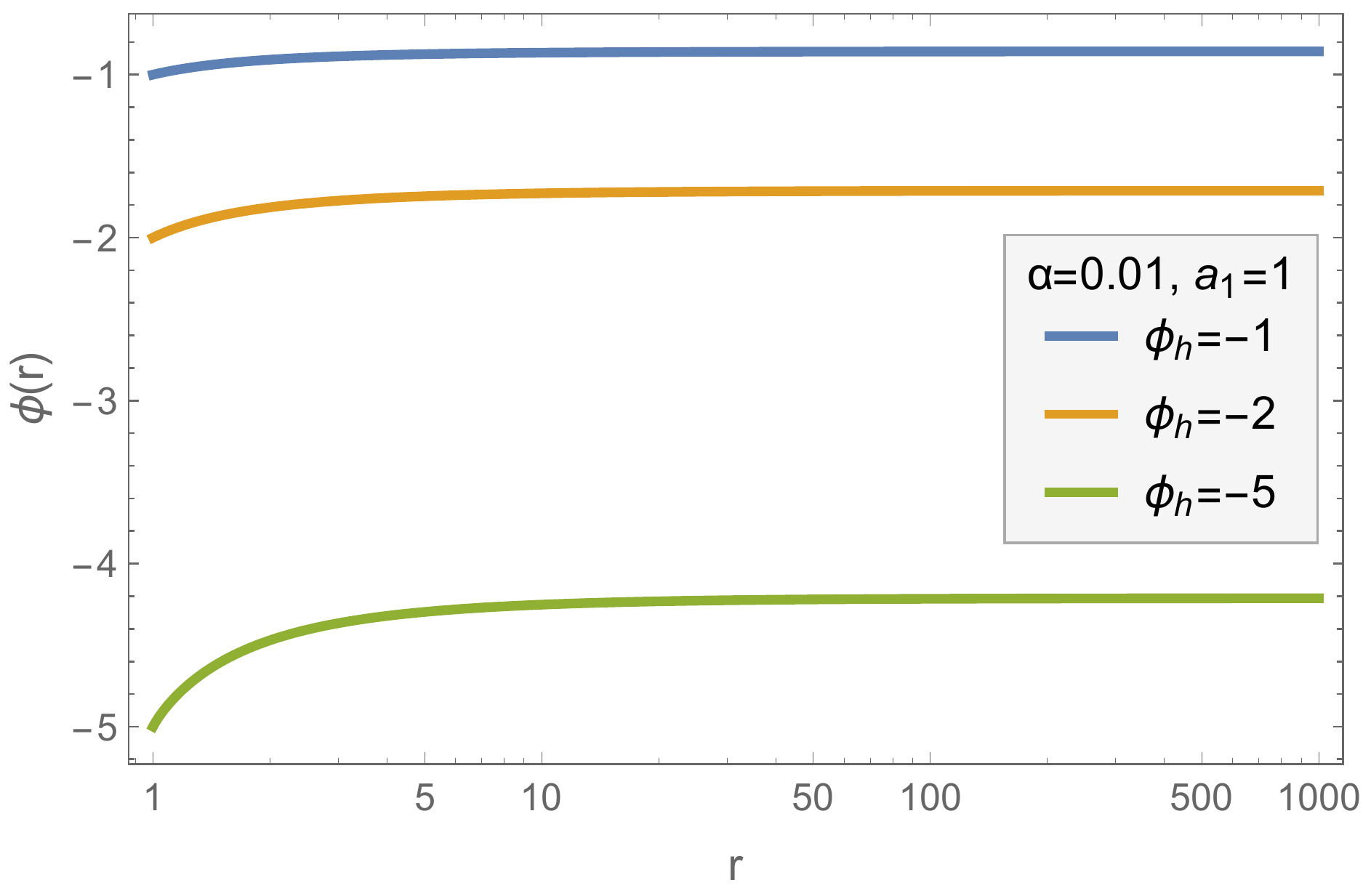}
\endminipage\hfill \hspace*{-2.0cm}
\minipage{0.52\textwidth}
  \includegraphics[width=\linewidth]{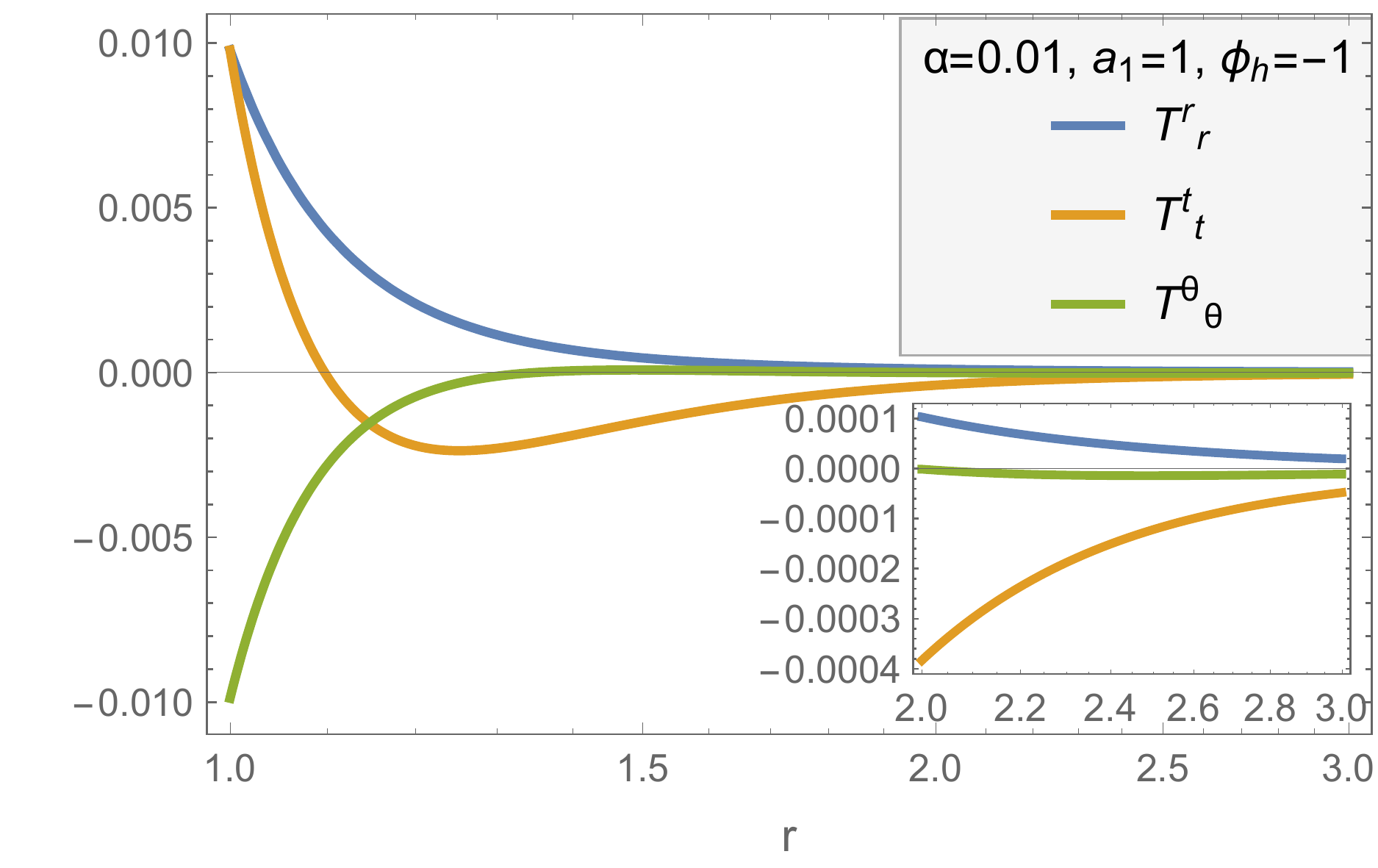}
\endminipage\hfill
 \caption{The scalar field $\phi$ (left plot), and the energy-momentum tensor
$T_{\mu\nu}$ (right plot) in terms of the radial coordinate $r$, for $f(\phi)=\alpha \phi^2$.}
\label{Phi_phi2}
\end{figure}

\begin{figure}[t!]
\minipage{0.49\textwidth}
  \includegraphics[width=\linewidth]{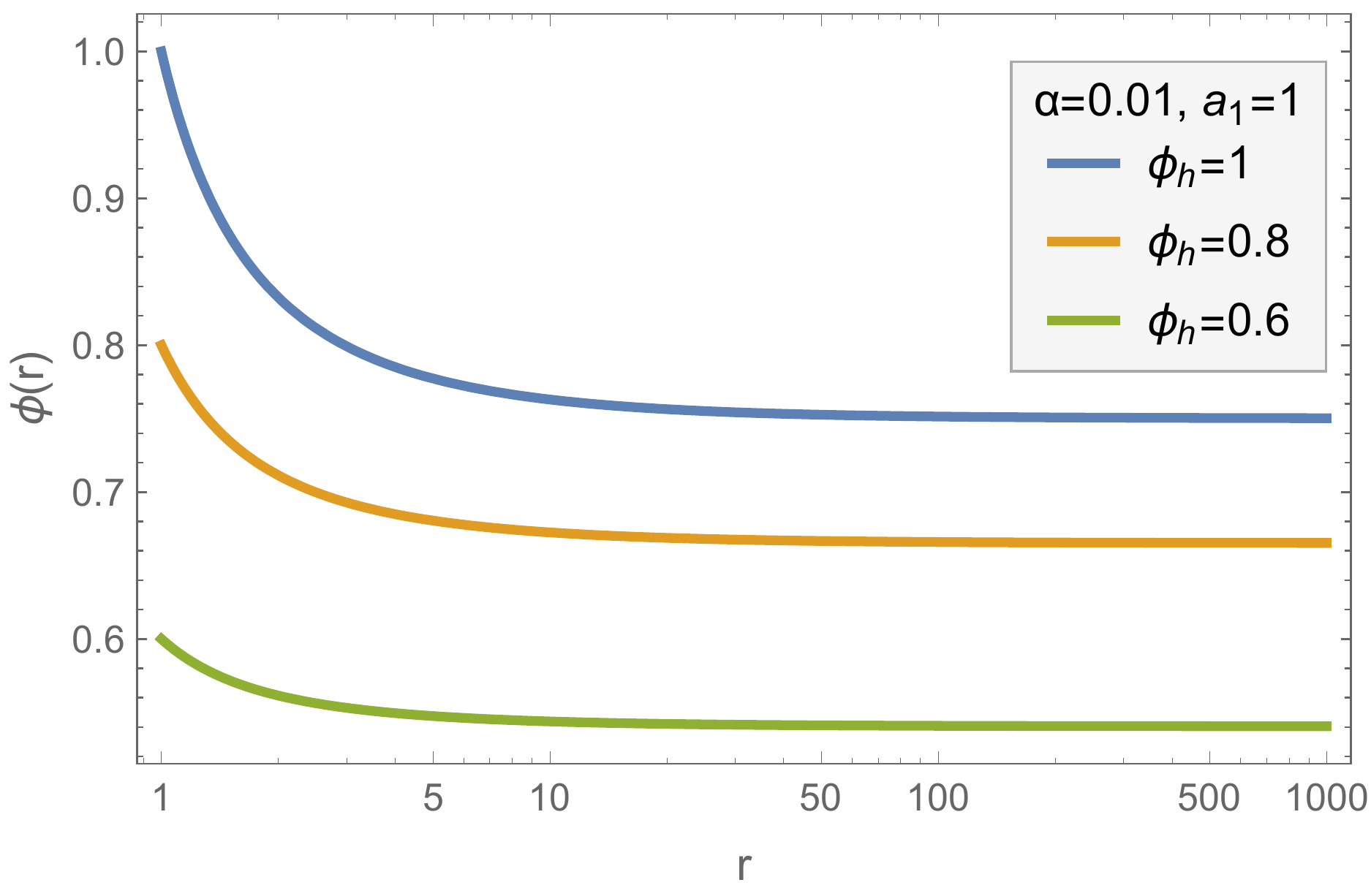}
\endminipage\hfill \hspace*{-2.0cm}
\minipage{0.52\textwidth}
  \includegraphics[width=\linewidth]{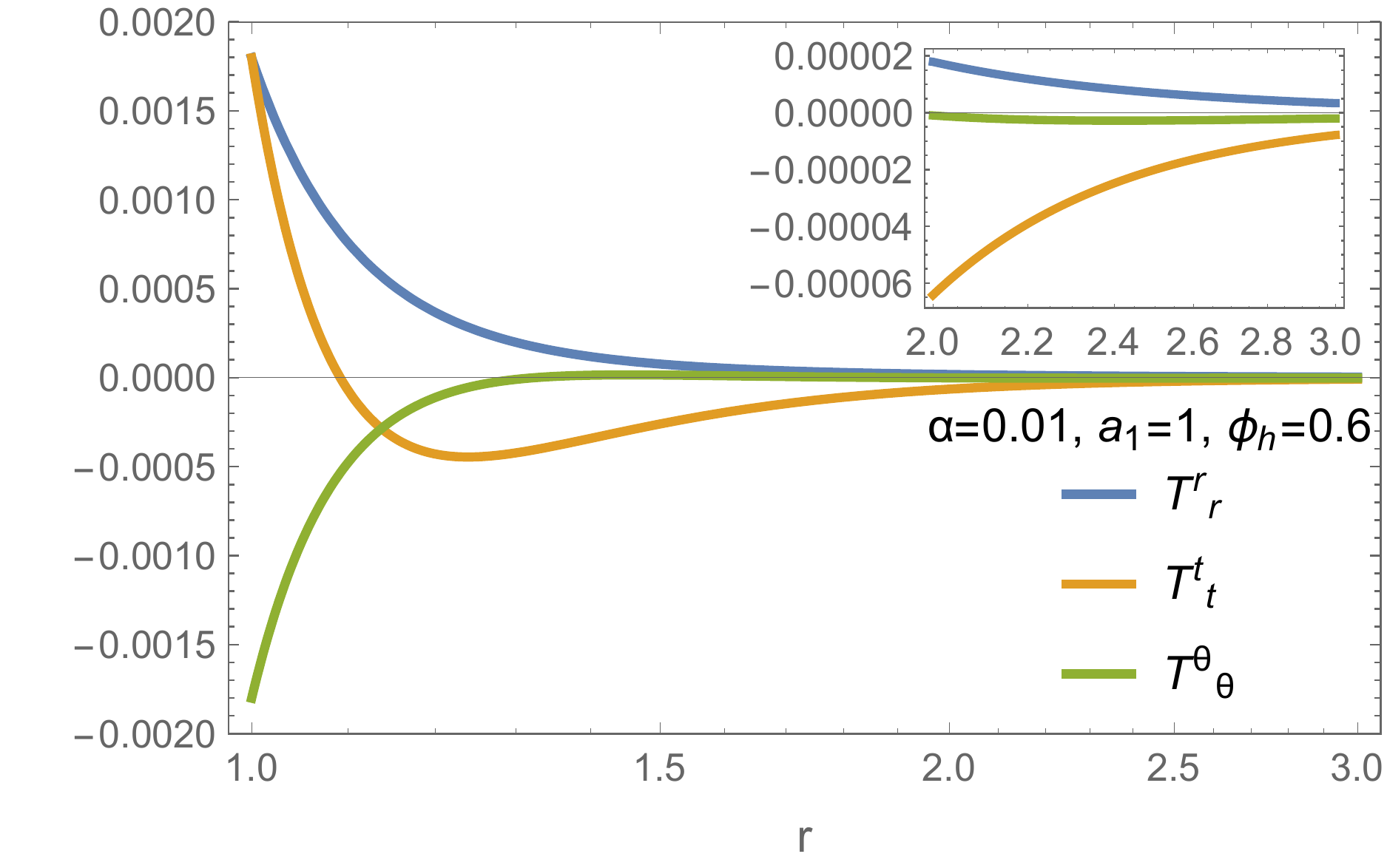}
\endminipage\hfill
\caption{The scalar field $\phi$ (left plot), and the energy-momentum tensor
$T_{\mu\nu}$ (right plot) in terms of the radial coordinate $r$, for $f(\phi)=\alpha \phi^4$.}
\label{Phi_phi4}
\end{figure}

Next, we consider the case where $f(\phi)=\alpha \phi^{2n}$, with $n \geq 1$. Since the
coupling function must be positive-definite, we assume again that $\alpha>0$. The first
constraint for the evasion of the novel no-hair theorem, $\dot f \phi' <0$ near the horizon,
now translates to $\phi_h \phi'_h<0$. Therefore, two classes of solutions appear for each value
of $n$: one for $\phi_h>0$, where $\phi'_h<0$ and the solution for the scalar field decreases
with $r$, and one for $\phi_h<0$, where $\phi'_h>0$ and the scalar field increases away
from the black-hole horizon. At the left plot of Fig. \ref{Phi_phi2}, we depict the first family
of solutions with $\phi_h<0$ and $\phi'_h>0$ for the choice $f(\phi)=\alpha \phi^2$ while,
at the left plot of Fig. \ref{Phi_phi4}, we depict the second class with $\phi_h>0$ and
$\phi'_h<0$ for the choice $f(\phi)=\alpha \phi^4$. The complimentary classes of solutions
may be easily derived in each case by reversing the signs of $\phi_h$ and $\phi'_h$.

The form of the energy-momentum tensor components for the two choices  $f(\phi)=\alpha \phi^2$
and  $f(\phi)=\alpha \phi^4$, and for two indicative solutions, are depicted at the right plots
of Figs. \ref{Phi_phi2} and \ref{Phi_phi4}, respectively. We observe that the qualitative
behaviour of the three components largely remain the same despite the change in the form
of the coupling function $f(\phi)$ (note also the resemblance with the behaviour
depicted at the right plot of Fig. \ref{Phi_Exp}). In fact, the asymptotic behaviour of
$T_{\mu\nu}$ near the black-hole horizon and radial infinity is fixed, according to 
Eqs. (\ref{Ttt-rh})-(\ref{Tthth-rh}) and (\ref{Tmn-far}), respectively. Independently
of the form of the coupling function $f(\phi)$, at asymptotic infinity $T^r_{\,\,r}$
approaches zero from the positive side, while $T^t_{\,\,t}$ and $T^{\theta}_{\,\,\theta}$
do the same from the negative side. In the near-horizon regime, 
Eqs. (\ref{Ttt-rh})-(\ref{Tthth-rh}) dictate that 
\begin{align}
\sign\left(T^t_{\;\,t}\right)_h,\;\sign\left(T^r_{\;\,r}\right)_h\sim & -\sign\left(\phi'_h\dot{f}_h\right),\\
\sign\left(T^\theta_{\;\,\theta}\right)_h\sim & +\sign\left(\phi'_h\dot{f}_h\right).
\end{align}
Using that $\dot f_h \phi'_h<0$ and the scaling behaviour of the metric functions near
the horizon, we may easily derive that  $(T^t_{\,\,t})_h$ and $(T^r_{\,\,r})_h$ always
assume positive values while ($T^\theta_{\,\,\theta})_h$ assumes a negative one.
The form of $f(\phi)$ merely changes the magnitude of these
asymptotic values: in the case of a polynomial coupling function, the higher the degree is,
the larger the asymptotic values near the horizon are.

\begin{figure}[b!]
\minipage{0.5\textwidth}
  \includegraphics[width=\linewidth]{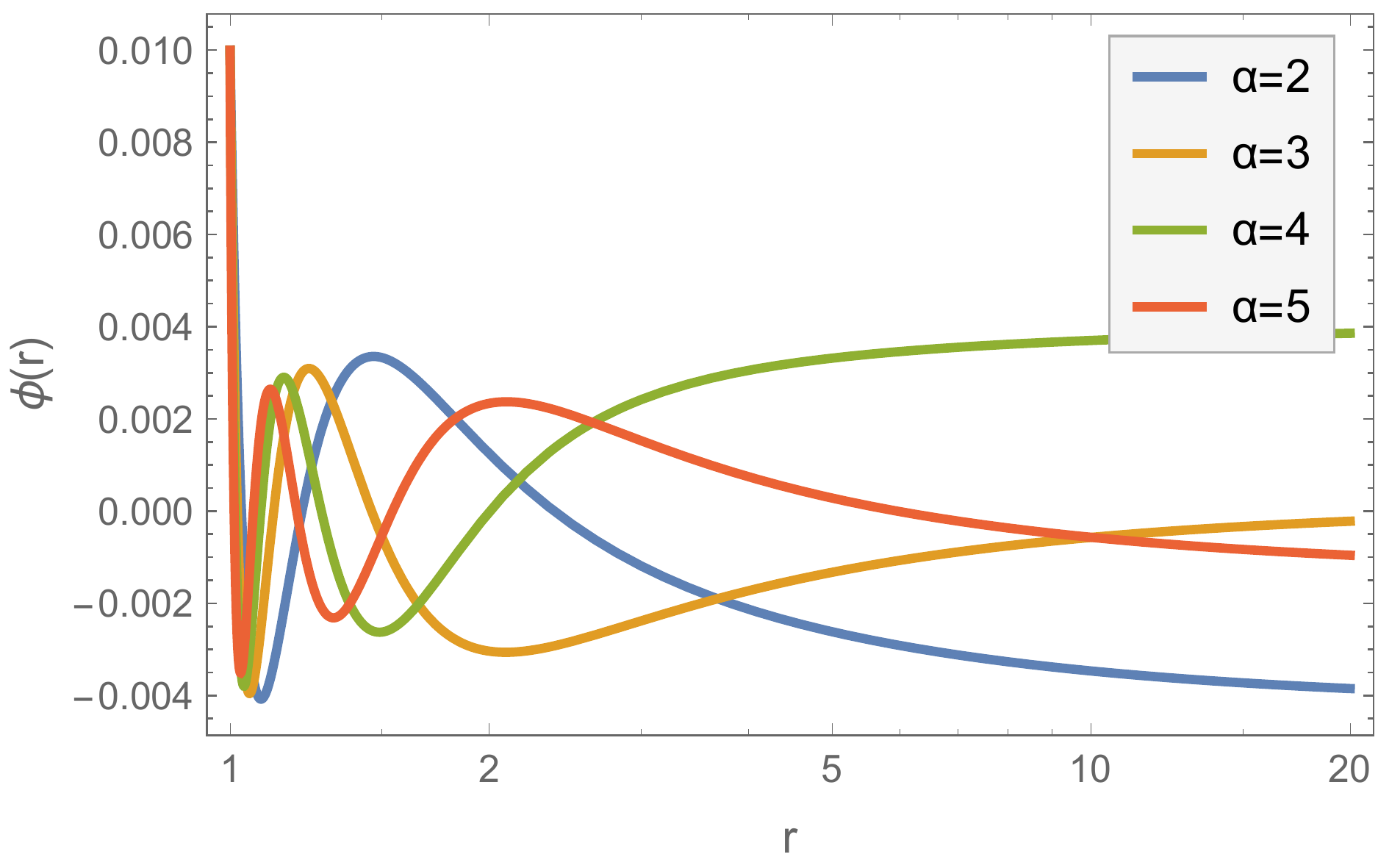}
\endminipage\hfill \hspace*{-2.0cm}
\minipage{0.5\textwidth}
  \includegraphics[width=\linewidth]{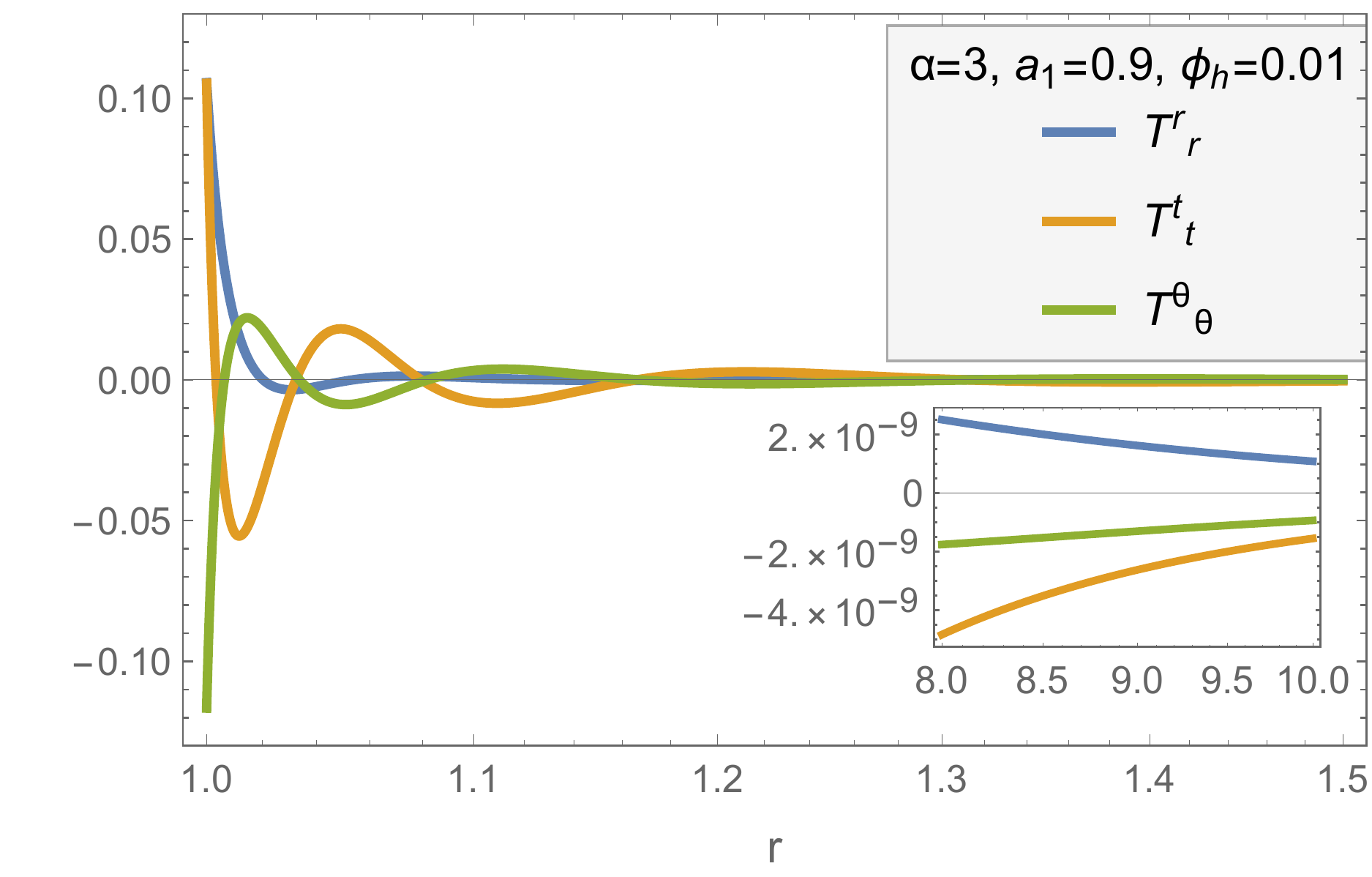}
\endminipage\hfill
\caption{The scalar field $\phi$ (left plot), and the energy-momentum tensor
$T_{\mu\nu}$ (right plot) in terms of the radial coordinate $r$, for $f(\phi)=\alpha \phi^2$
and various values of the coupling constant $\alpha$.}
\label{Phi_alpha_phi2}
\end{figure}

One could assume that the intermediate behaviour of the scalar field and the energy-momentum
tensor always remains qualitatively the same. In fact, this is not so. Let us fix for simplicity
the values of $r_h$ and $\phi_h$, and gradually increase the value of the coupling
parameter $\alpha$; this has the effect of increasing the magnitude of the GB source-term
appearing in the equation of motion (\ref{phi-eq_0}) for $\phi$. As an indicative case,
at the left plot of Fig. \ref{Phi_alpha_phi2}, we
depict the behaviour of $\phi$ for four large values of $\alpha$, and, at the right
plot, the $T_{\mu\nu}$ for one of these solutions. We observe that all of these quantities
are not monotonic any more; they go through a number of maxima or minima -- with
that number increasing with the value of $\alpha$ -- before reaching their asymptotic
values at infinity. Note that the near-horizon behaviour of both $\phi$ and $T_{\mu\nu}$
is still the one that guarantees the evasion of the no-hair theorem. We may thus
conclude that the presence of the GB term in the theory not only ensures that
the asymptotic solutions (\ref{A-rh})-(\ref{phi-rh}) and (\ref{Afar})-(\ref{phifar}) may be
smoothly connected to create a regular black hole but it allows for this to happen
even in a non-monotonic way.

\begin{figure}[t!]
\minipage{0.50\textwidth}
  \includegraphics[width=\linewidth]{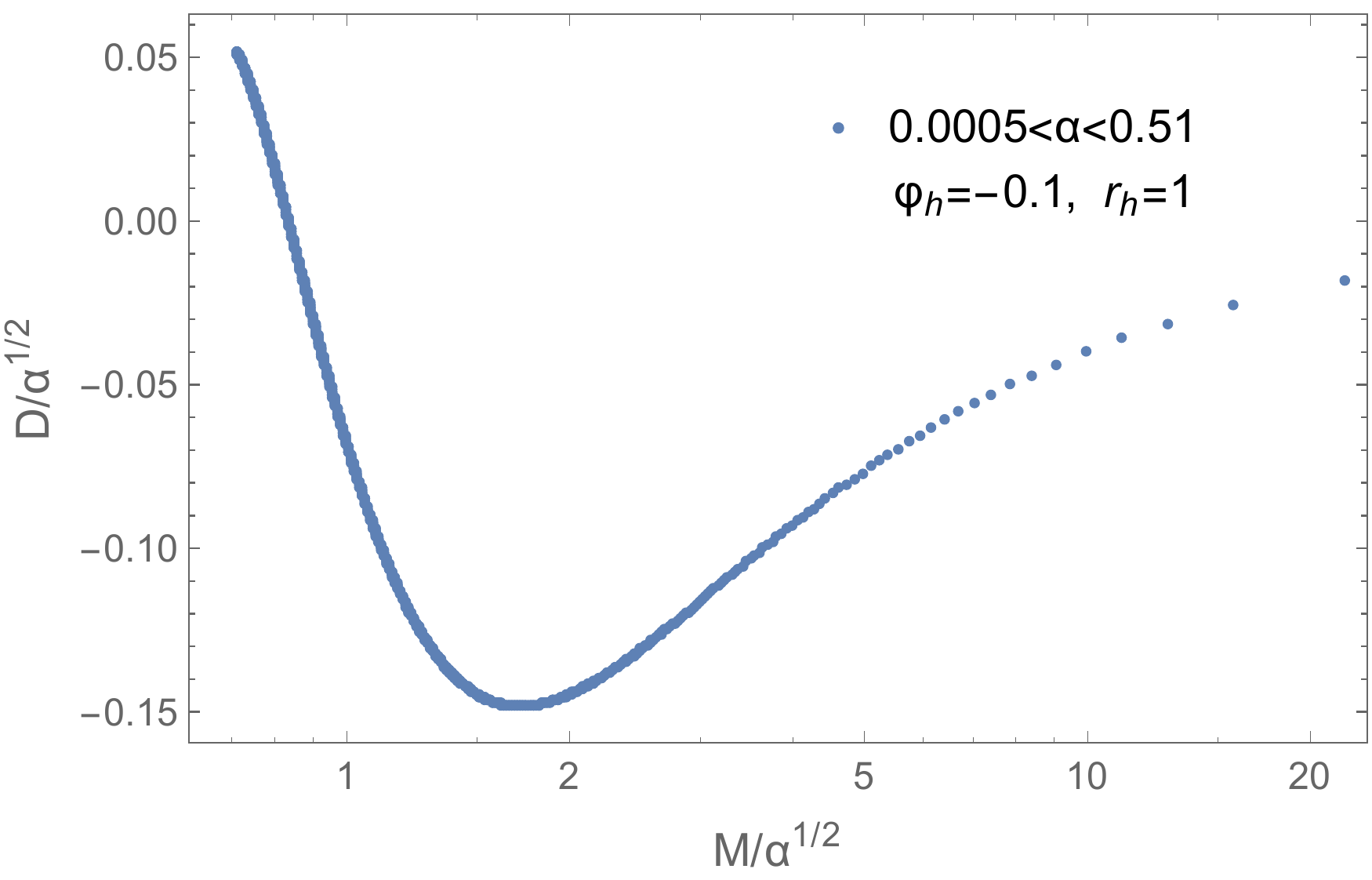}
\endminipage\hfill \hspace*{-2.0cm}
\minipage{0.51\textwidth}
  \includegraphics[width=\linewidth]{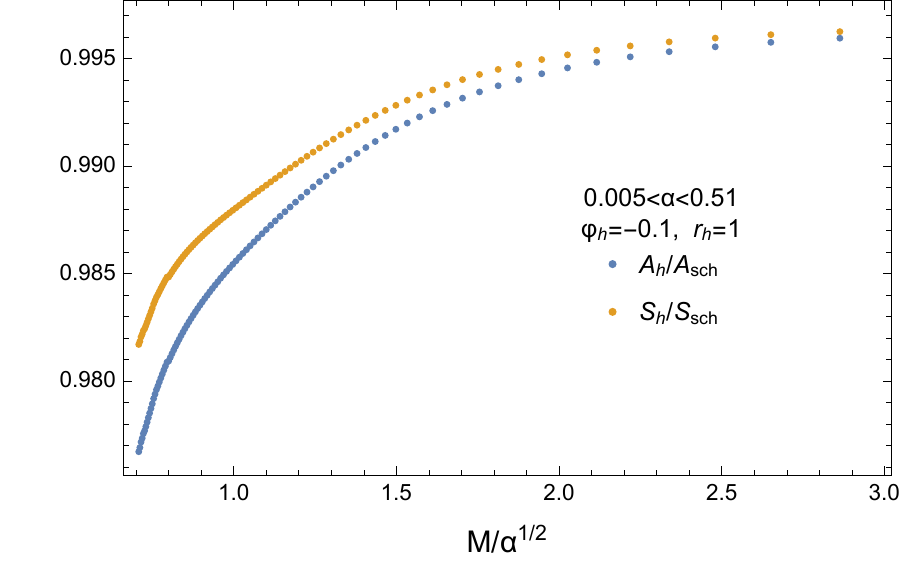}
\endminipage\hfill
\caption{The scalar charge $D$ (left plot), and the ratios $A_h/A_{Sch}$ and $S_h/S_{Sch}$ 
(right plot) in terms of the mass  $M$, for $f(\phi)=\alpha \phi^2$.}
\label{D-AS-even}
\end{figure}

Let us also study the characteristics of this class of black-hole solutions arising for an
even polynomial coupling function. In Fig. \ref{D-AS-even} (left plot), we depict the
scalar charge $D$ in terms of the mass $M$ of the black hole, for the quadratic coupling 
function $f(\phi)=\alpha \phi^2$: we observe that, in this case, the function $D(M)$ is
not monotonic in the small-mass regime but it tends again to zero for large values of
its mass. In terms of the near-horizon value $\phi_h$,
the scalar charge exhibits the expected behaviour: for large values of $\phi_h$, the
effect of the GB term becomes important and $D$ increases; on the other hand, for
vanishing $\phi_h$, i.e. a vanishing coupling function, the scalar charge also vanishes
-- in order to minimise the number of figures, we refrain from showing plots depicting
the anticipated behaviour; in the same spirit, we present no new plots for the quartic
coupling function as it leads to exactly the same qualitative behaviour. 

Turning to the horizon area $A_h$ and entropy $S_h$ of these black-hole solutions, we find
that, in terms of the mass $M$, they both quickly increase, showing a profile similar to
that of Fig. \ref{AS-exp} (left plot) for the exponential case. The ratio $A_h/A_{Sch}$
remains again below unity over the whole mass regime, and interpolates between a
value corresponding to the lowest allowed value of the mass, according to Eq. (\ref{con-f}),
and the asymptotic Schwarzschild value at the large-mass limit. The entropy ratio
$S_h/S_{Sch}$, on the other hand, is found to have a different profile by remaining
now always below unity - this feature points perhaps towards a thermodynamic instability
of the `even polynomial' GB black holes compared to the Schwarzschild solution.


\subsection{Odd Polynomial Function}

\begin{figure}[t!]
\minipage{0.49\textwidth}
  \includegraphics[width=\linewidth]{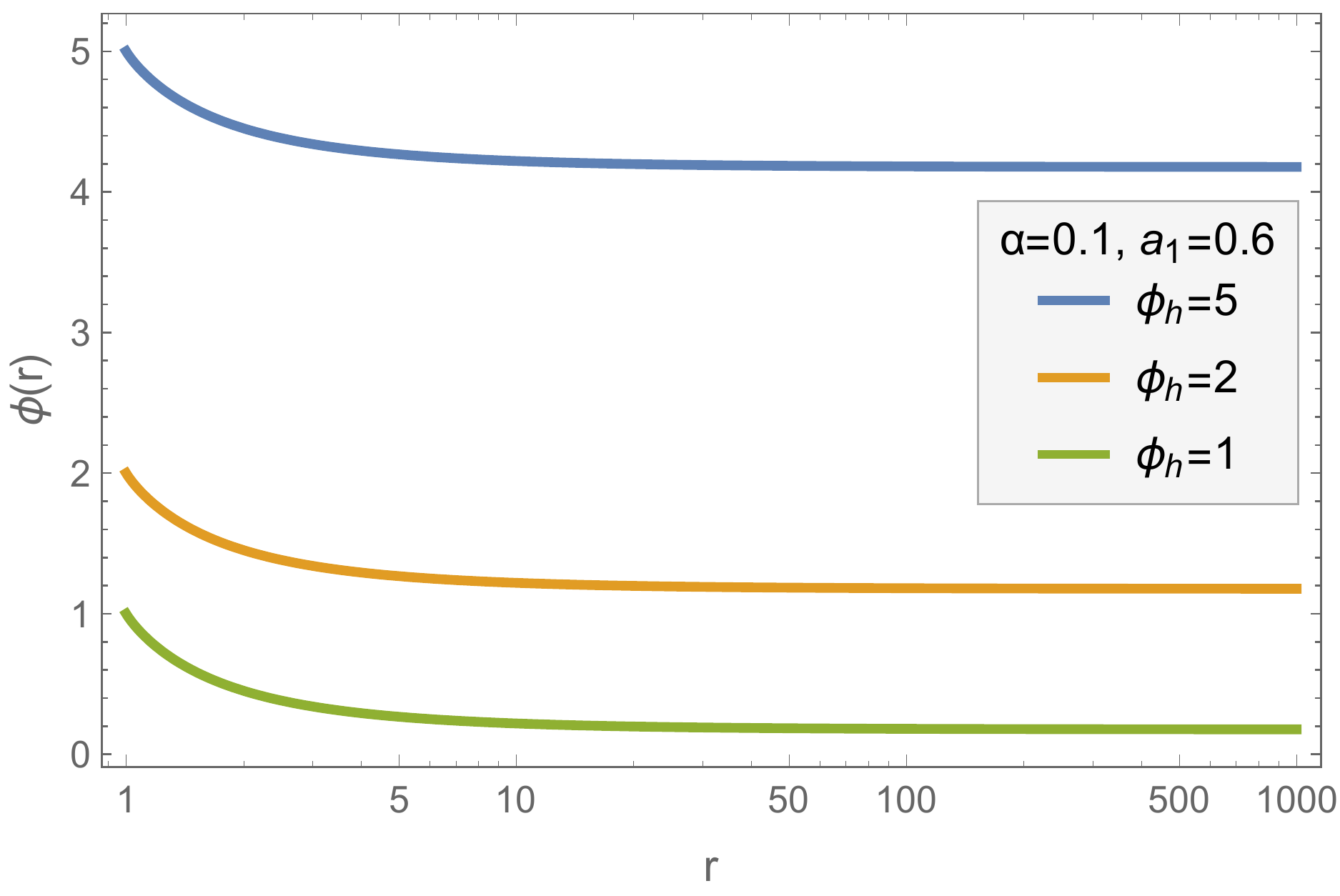}
\endminipage\hfill \hspace*{-1.5cm}
\minipage{0.5\textwidth}
  \includegraphics[width=\linewidth]{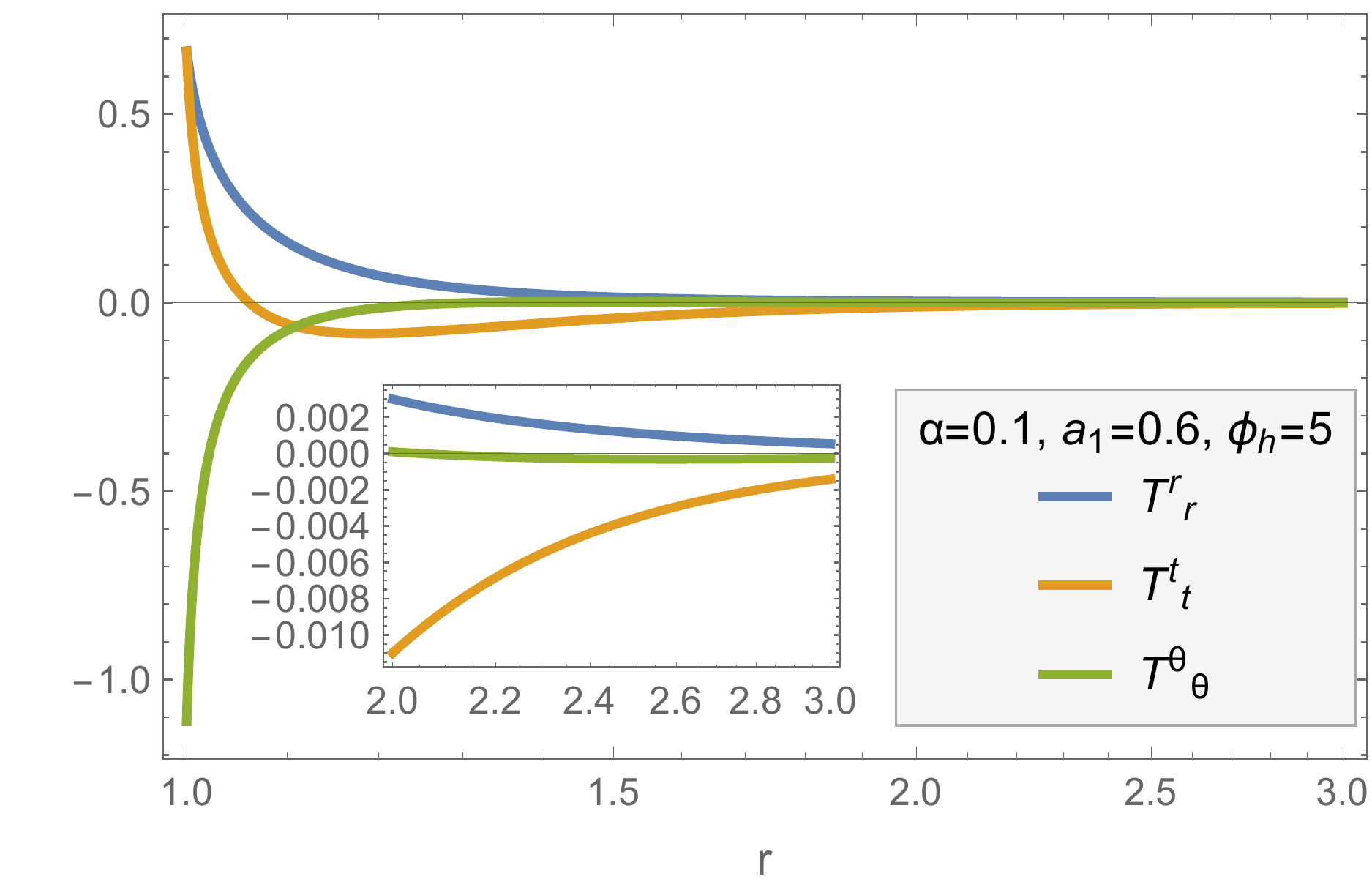}
\endminipage\hfill
\caption{The scalar field $\phi$ (left plot), and the energy-momentum tensor
$T_{\mu\nu}$ (right plot) in terms of the radial coordinate $r$, for $f(\phi)=\alpha \phi$.}
\label{Phi_phi}
\end{figure}

We now consider the case of $f(\phi)=\alpha\,\phi^{2n+1}$, with $n \geq 0$
and $\alpha>0$. Here, the constraint $\dot f \phi'<0$ translates to 
$\phi^{2n} \phi' <0$, or simply to $\phi'_h<0$ for all solutions. 
In Fig. \ref{Phi_phi}, we have chosen the linear case, i.e. $f(\phi)=\alpha \phi$,
and presented an indicative family of solutions for the scalar field (left plot)
and the components of the energy-momentum tensor for one of them (right plot).
The decreasing profile of $\phi $ for all solutions, as we move away from the
black-hole horizon, is evident and in agreement with the above constraint.
The energy-momentum tensor clearly satisfies the analytically predicted behaviour
at the two asymptotic regimes, that once again ensures the evasion of the novel
no-hair theorem.

\begin{figure}[b!]
\minipage{0.49\textwidth}
  \includegraphics[width=\linewidth]{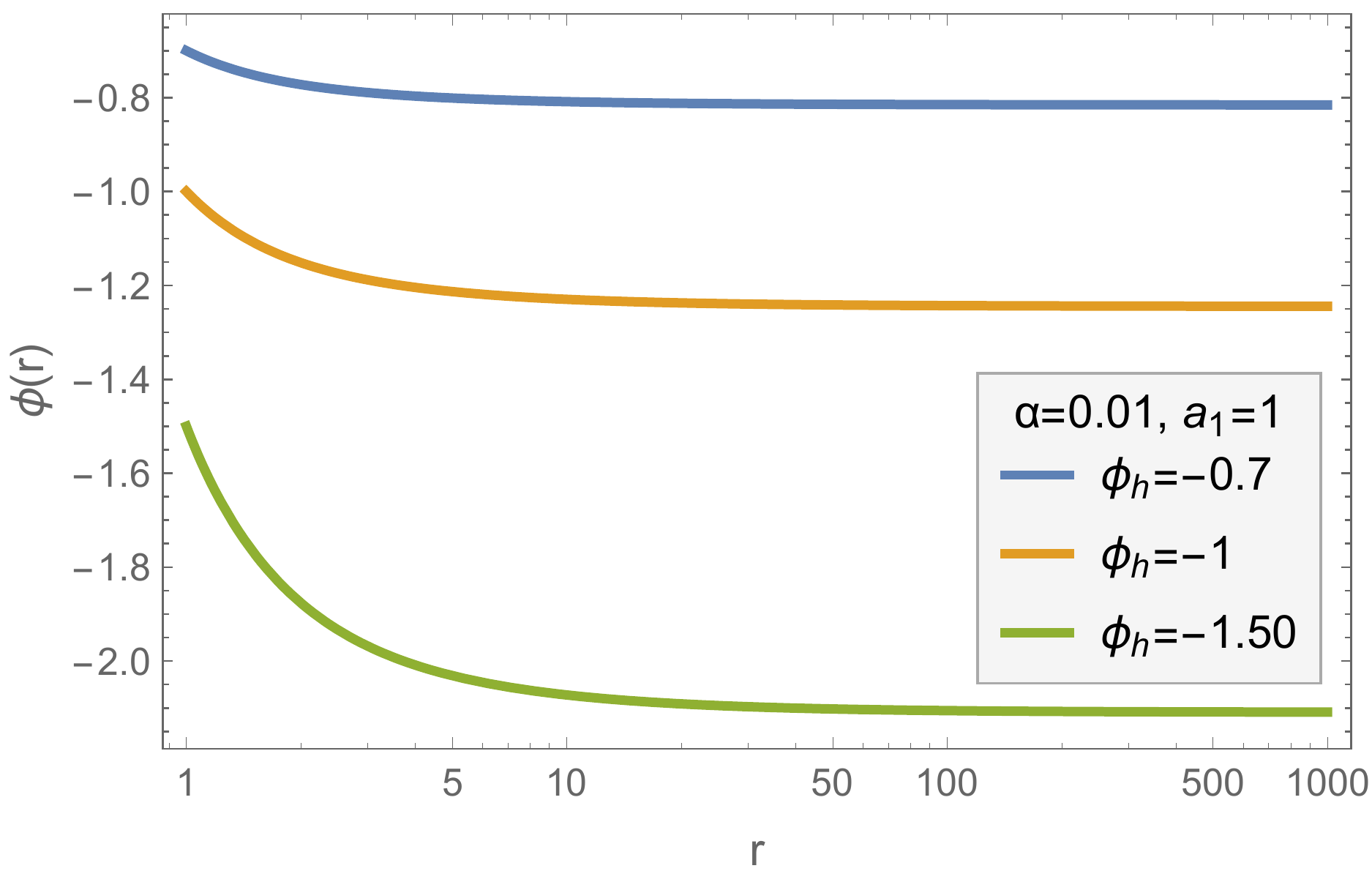}
\endminipage\hfill \hspace*{-1.5cm}
\minipage{0.51\textwidth}
  \includegraphics[width=\linewidth]{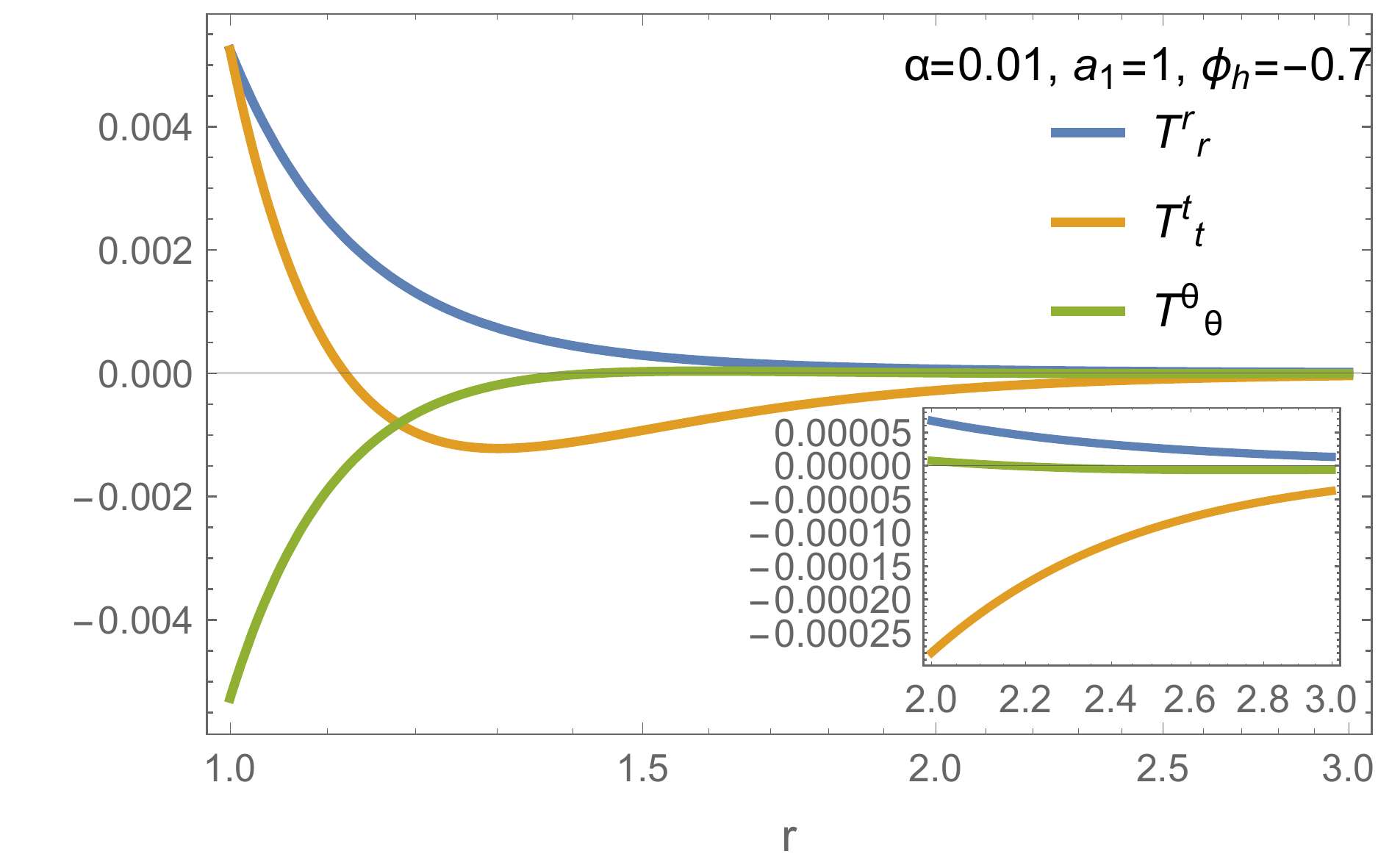}
\endminipage\hfill
\caption{The scalar field $\phi$ (left plot), and the energy-momentum tensor
$T_{\mu\nu}$ (right plot) in terms of the radial coordinate $r$, for $f(\phi)=\alpha \phi^3$.}
\label{Phi_phi3}
\end{figure}

The two plots in Fig. \ref{Phi_phi3}, left and right, depict the
same quantities but for the case $f(\phi)=\alpha \phi^3$. Their profile
agrees with that expected for a regular, black-hole solution with a scalar
hair. The alerted reader may notice that, here, we have chosen to present
solutions with $\phi_h<0$; for an odd polynomial coupling function, these
should have been prohibited under the constraint $f(\phi)>0$, that follows from
the old no-hair theorem \cite{NH-scalar, ABK1}. Nevertheless, regular
black-hole solutions with a
non-trivial scalar field do emerge, that do not seem to satisfy $f(\phi)>0$.
A set of such solutions are shown at the left plot of Fig. \ref{Phi_phi3} (we
refrain from showing the set of solutions with $\phi_h>0$ as these have
similar characteristics). As we observe, all of them
obey the $\phi'_h<0$ constraint imposed by the evasion of the novel
no-hair theorem, and lead to the expected behaviour of $T_{\mu\nu}$;
the latter may be clearly seen at the right plot of Fig. \ref{Phi_phi3} where
such a ``prohibited'' solution is plotted. The behaviour of the metric
components and the GB term continue to be given by plots similar to
the ones in Fig. \ref{Metric_GB}. 

The emergence of solutions that violate the constraint $f(\phi)>0$ is in
fact a general feature of our analysis and not an isolated finding in the
case of odd polynomial coupling function. Apparently, the presence of the
coupling of the scalar field to the GB term not only opens the way for 
black-hole solutions to emerge but renders the old no-hair theorem
incapable of dictating when this may happen. Looking more carefully at
the argument on which the old no-hair theorem was based on
\cite{NH-scalar, ABK1}, one readily
realizes that this involves the integral of the scalar equation over the
entire exterior regime, and thus the global solution of the field equations
whose characteristics cannot be predicted beforehand. In contrast, the
evasion of the novel no-hair theorem is based on local quantities, such
as the energy-momentum components and their derivatives at particular
radial regimes, which may be easily computed. In addition, it is indifferent
to the behaviour of the solution in the intermediate regime, which may
indeed exhibit an arbitrary profile as the one presented in the plots of
Fig. \ref{Phi_alpha_phi2}. In fact, all solutions found
in the context of our analysis, with no exception, satisfy the constraints 
that ensure the evasion of the novel no-hair theorem.

\begin{figure}[t!]
\minipage{0.49\textwidth}
  \includegraphics[width=\linewidth]{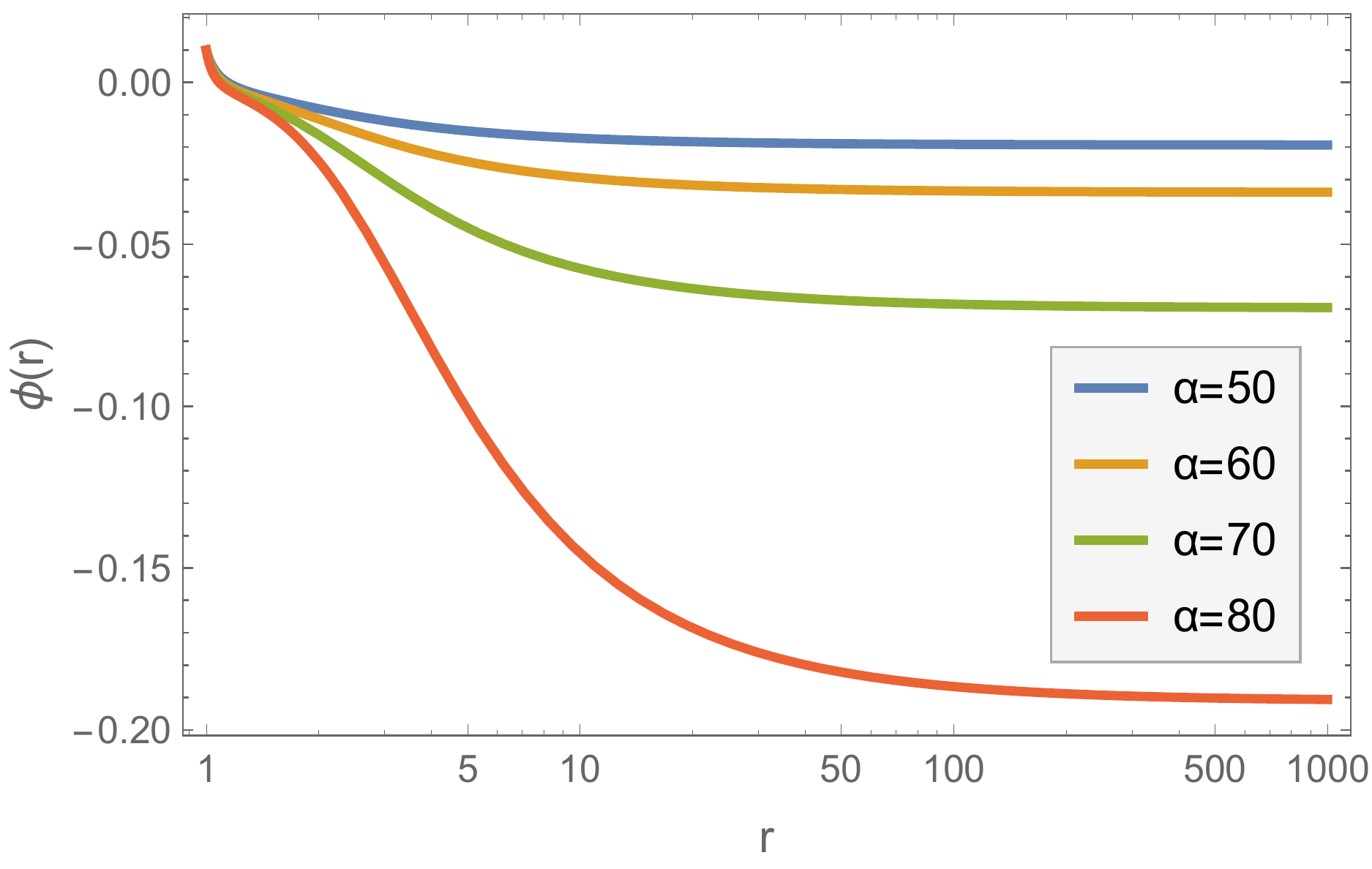}
\endminipage\hfill \hspace*{-1.5cm}
\minipage{0.49\textwidth}
  \includegraphics[width=\linewidth]{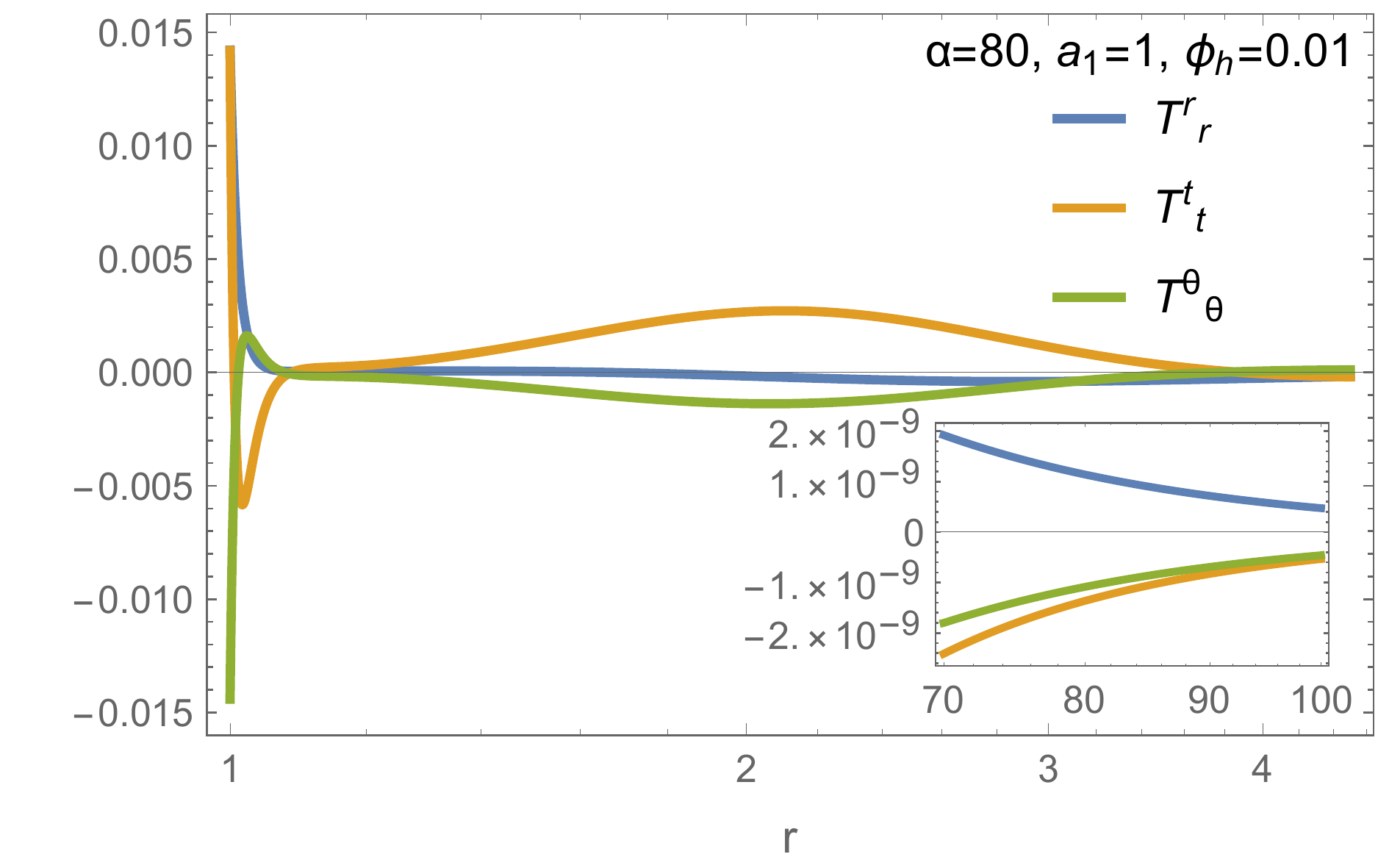}
\endminipage\hfill
\caption{The scalar field $\phi$ (left plot), and the energy-momentum tensor
$T_{\mu\nu}$ (right plot) in terms of the radial coordinate $r$, for $f(\phi)=\alpha \phi^3$
and a variety of values of the coupling constant $\alpha$.}
\label{Phi_alpha_phi3}
\end{figure}

In this case, too, one may derive solutions for a variety of values of the
coupling constant $\alpha$, as long as these obey the constraint (\ref{con-f}).
In Fig. \ref{Phi_alpha_phi3} (left plot), we depict a family of solutions with $\phi_h=0.01$
and a variety of values of $\alpha$. This family of solutions present a less
monotonic profile compared to the one exhibited by the solutions in Fig. \ref{Phi_phi3}.
The components of the energy-momentum tensor for one of these solutions is
depicted at the right plot of Fig. \ref{Phi_alpha_phi3}, and they present a more evolved
profile with the emergence of minima and maxima between the black-hole horizon and
radial infinity. We also notice that the solutions for the scalar field, although they start
from the positive-value regime ($\phi_h=0.01$), they cross to negative values for fairly
small values of the radial coordinate. This behaviour causes the odd coupling function
to change sign along the radial regime, a feature that makes any application of the
old no-hair theorem even more challenging. 

\begin{figure}[t!]
\minipage{0.50\textwidth}
  \includegraphics[width=\linewidth]{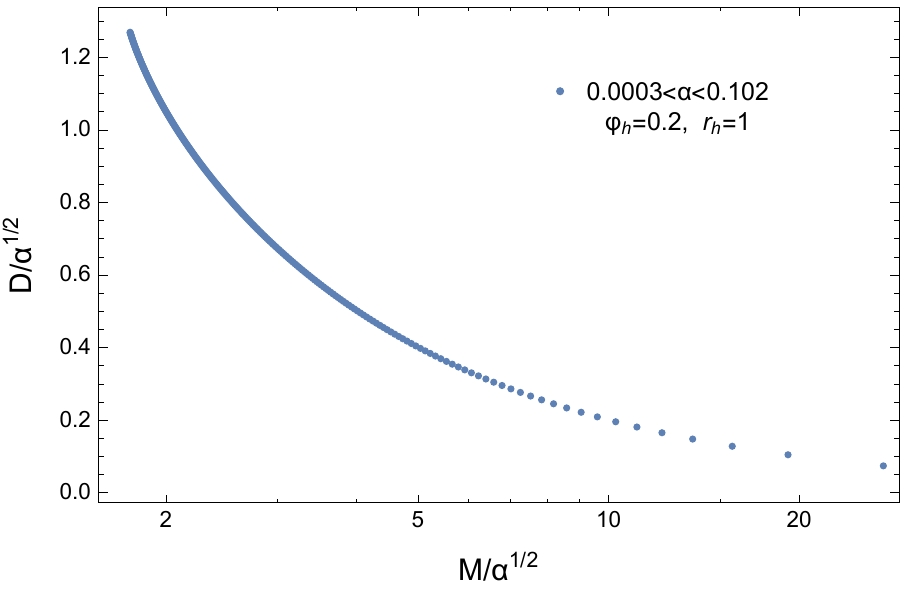}
\endminipage\hfill \hspace*{-2.0cm}
\minipage{0.51\textwidth}
  \includegraphics[width=\linewidth]{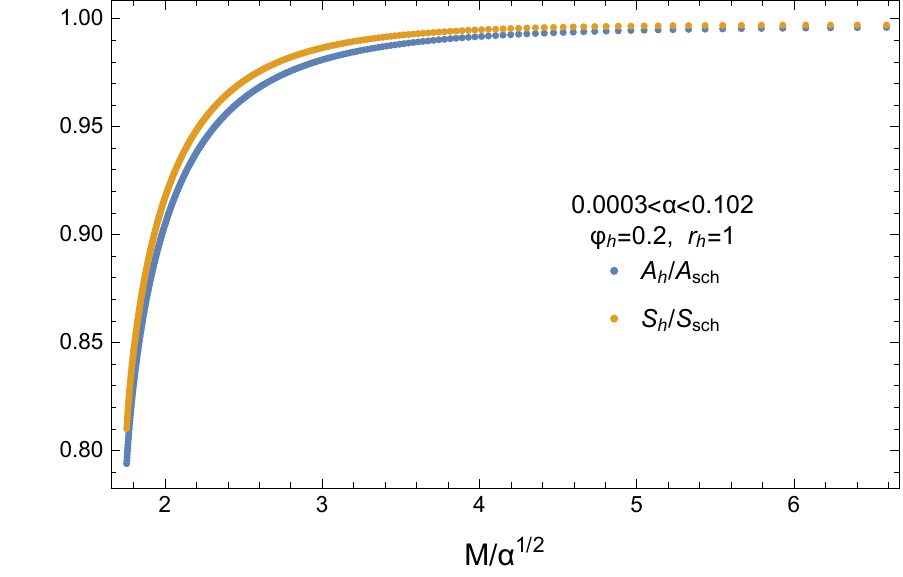}
\endminipage\hfill
\caption{The scalar charge $D$ (left plot), and the ratios $A_h/A_{Sch}$ and $S_h/S_{Sch}$ 
(right plot) in terms of the mass  $M$, for $f(\phi)=\alpha \phi$.}
\label{D-AS-linear}
\end{figure}

Turning again to the characteristics of the `odd polynomial' black-hole solutions, in
Fig. \ref{D-AS-linear} (left plot), we depict the scalar charge $D$ in terms of the
mass $M$ of the black hole, for the linear coupling function $f(\phi)=\alpha \phi$:
here, the function $D(M)$ is monotonic and approaches a vanishing asymptotic
value as $M$ increases. The scalar charge $D$ has no dependence on the initial
scalar-field value $\phi_h$ since it is the first derivative $f'(\phi)$ that appears
in the scalar equation (\ref{phi-eq_0}) and, for a linear function, this is merely a
constant. The horizon area $A_h$ and entropy $S_h$ exhibit again an increasing profile
in terms of $M$ similar to
that of Fig. \ref{AS-exp} (left plot) for the exponential case. The more informative
ratios $A_h/A_{Sch}$ and $S_h/S_{Sch}$ are given in the right plot of Fig. \ref{D-AS-linear}:
as in the case of quadratic and quartic coupling functions, both quantities remain
smaller than unity and interpolate between a lowest value corresponding to the
black-hole solution with the lowest mass and the Schwarzschild limit acquired at
the large mass limit.

\begin{figure}[t!]
\minipage{0.50\textwidth}
  \includegraphics[width=\linewidth]{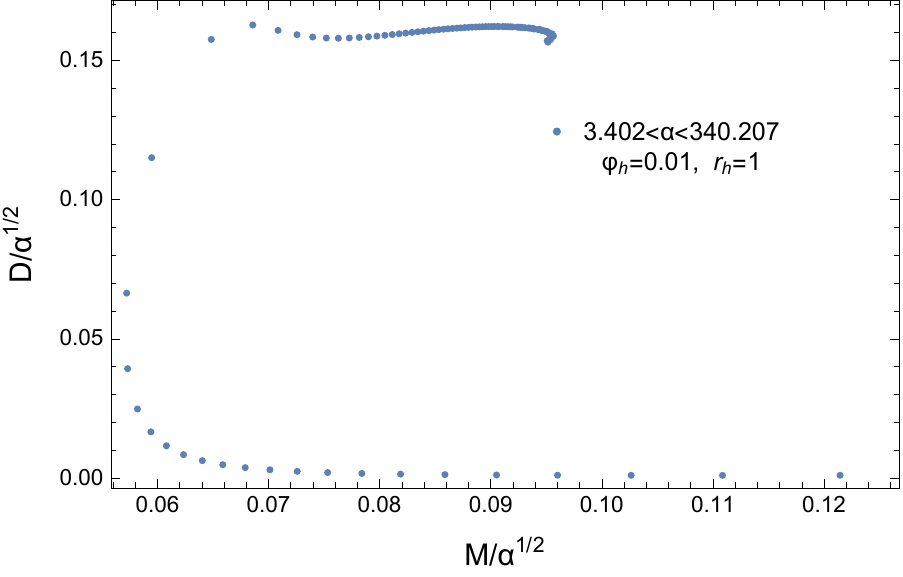}
\endminipage\hfill \hspace*{-2.0cm}
\minipage{0.51\textwidth}
  \includegraphics[width=\linewidth]{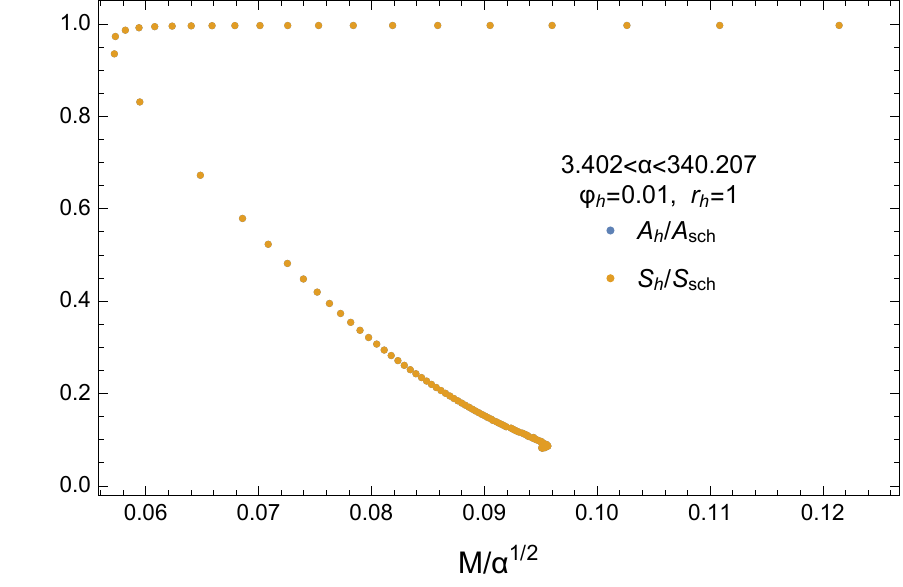}
\endminipage\hfill
\caption{The scalar charge $D$ (left plot), and the ratios $A_h/A_{Sch}$ and $S_h/S_{Sch}$ 
(right plot) in terms of the mass  $M$, for $f(\phi)=\alpha \phi^3$.}
\label{D-AS-qubic}
\end{figure}

We now address separately the characteristics of the black-hole solutions arising
in the case of the cubic coupling function $f(\phi)=\alpha \phi^3$ since here we
find a distinctly different behaviour. As mentioned above, also in this case, as
the coupling constant $\alpha$ increases, from zero to its maximum value (for
$r_h$ and $\phi_h$ fixed), solutions with no monotonic profile in terms of the
radial coordinate arise (see left plot of Fig. \ref{Phi_alpha_phi3}). We depict
the scalar charge $D$ in terms of the mass $M$ of the black hole, for the whole
$\alpha$-regime, in the left plot of Fig. \ref{D-AS-qubic}: we may easily observe
the emergence of two different branches of solutions (with a third, short one 
appearing at the end of the upper branch) corresponding to the same mass $M$. 
These branches appear at the small-mass limit of the solutions whereas for
large masses only one branch survives with a very small scalar charge. In
the right plot of Fig. \ref{D-AS-qubic}, we show the ratio $S_h/S_{Sch}$ in terms
of the mass $M$: this quantity also displays the existence of three branches 
with the one that is smoothly connected to the Schwarzschild limit having the
higher entropy. The two additional branches with the larger values of scalar
charge, compared to the one of the `Schwarzschild' branch, have a lower entropy
and they are probably less thermodynamically stable. This behaviour was not
observed in the case of the quadratic coupling function where more evolved
solutions for the scalar field also appeared (see left plot of Fig. \ref{Phi_phi2}):
there, the function $D(M)$ was not monotonic but was always single-valued; that
created short, disconnected  `branches' of solutions with slightly different
values of entropy ratio $S_h/S_{Sch}$ but all lying below unity. Let us finally
note that the horizon area ratio $A_h/A_{Sch}$, not shown here for brevity,
has the same profile as the one displayed in Fig. \ref{D-AS-linear} for the
linear function while the $D(\phi_h)$ function shows the anticipated increasing
profile as $\phi_h$, and thus the GB coupling, increases.

\subsection{Inverse Polynomial Function}

The next case to consider is the one where $f(\phi)=\alpha \phi^{-k}$, where
$k>0$, and $\alpha$ is also assumed to be positive, for simplicity. Let us 
consider directly some indicative cases:

\begin{figure}[b!]
\minipage{0.52\textwidth}
  \includegraphics[width=\linewidth]{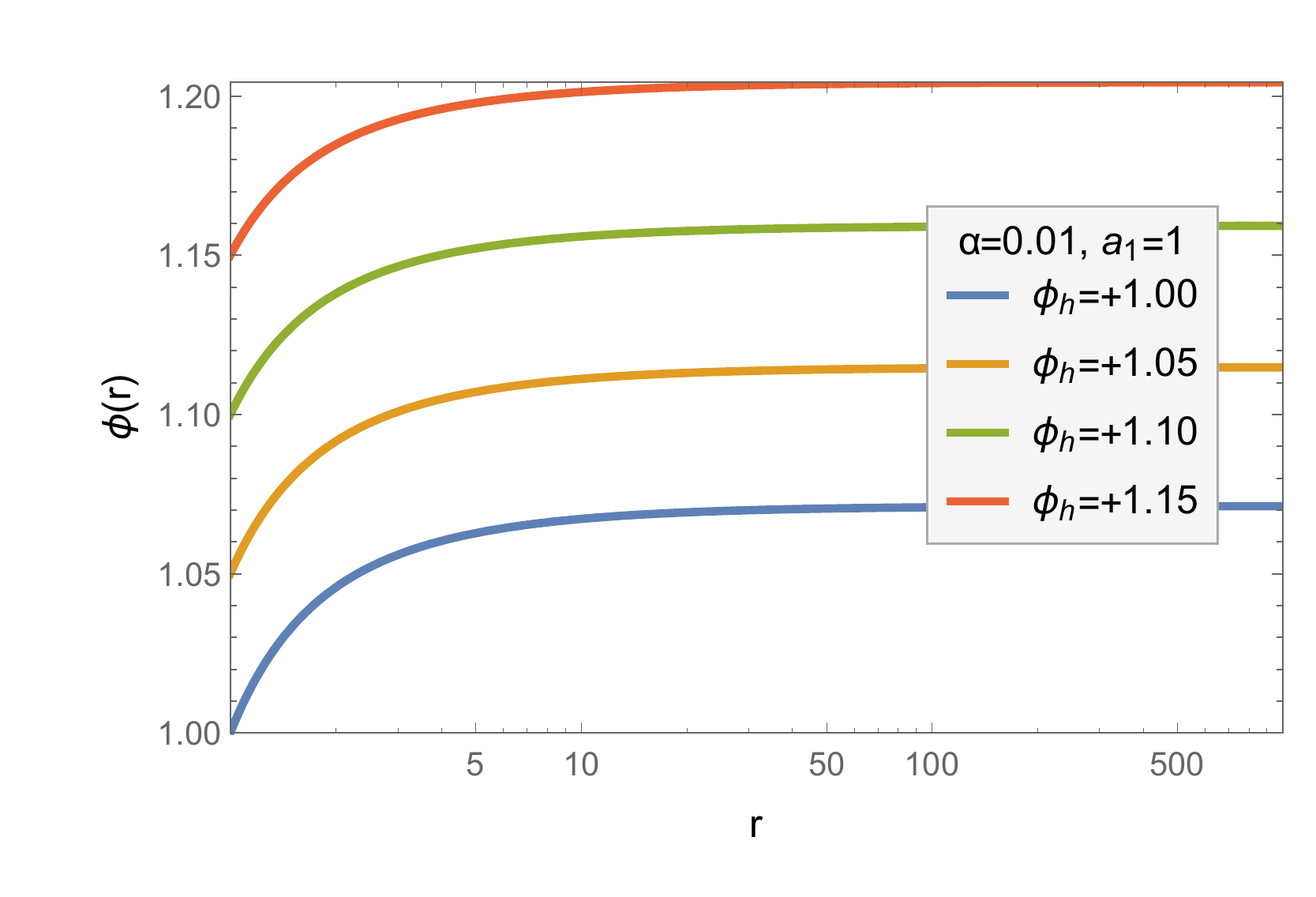}
\endminipage\hfill \hspace*{-1.8cm}
\minipage{0.52\textwidth}
  \includegraphics[width=\linewidth]{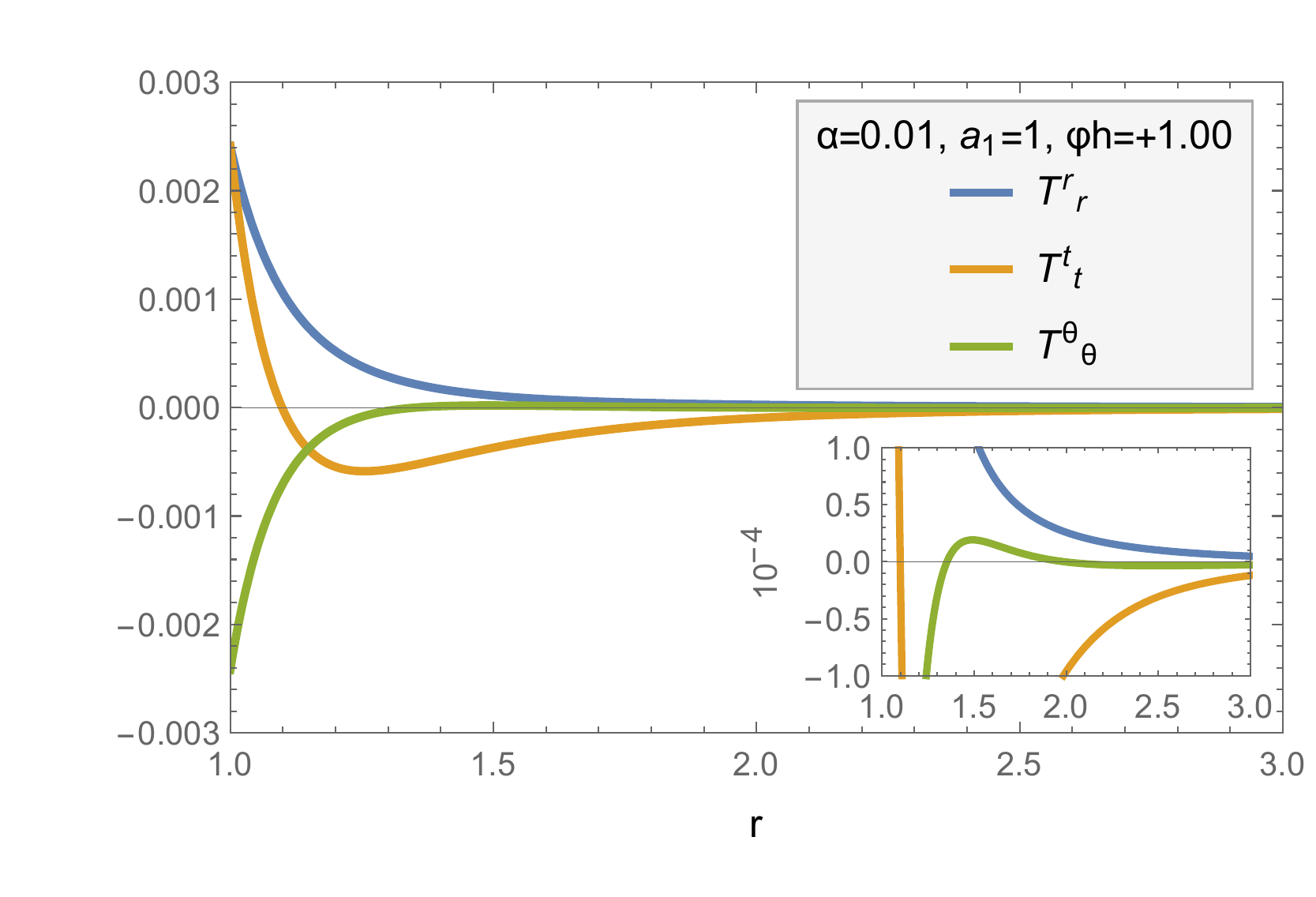}
\endminipage\hfill
\caption{The scalar field $\phi$ (left plot), and the energy-momentum tensor
$T_{\mu\nu}$ (right plot) in terms of the radial coordinate $r$, for $f(\phi)=\alpha/\phi$.}
\label{Phi_neg1}
\end{figure}

\begin{figure}[t!]
\minipage{0.52\textwidth}
  \includegraphics[width=\linewidth]{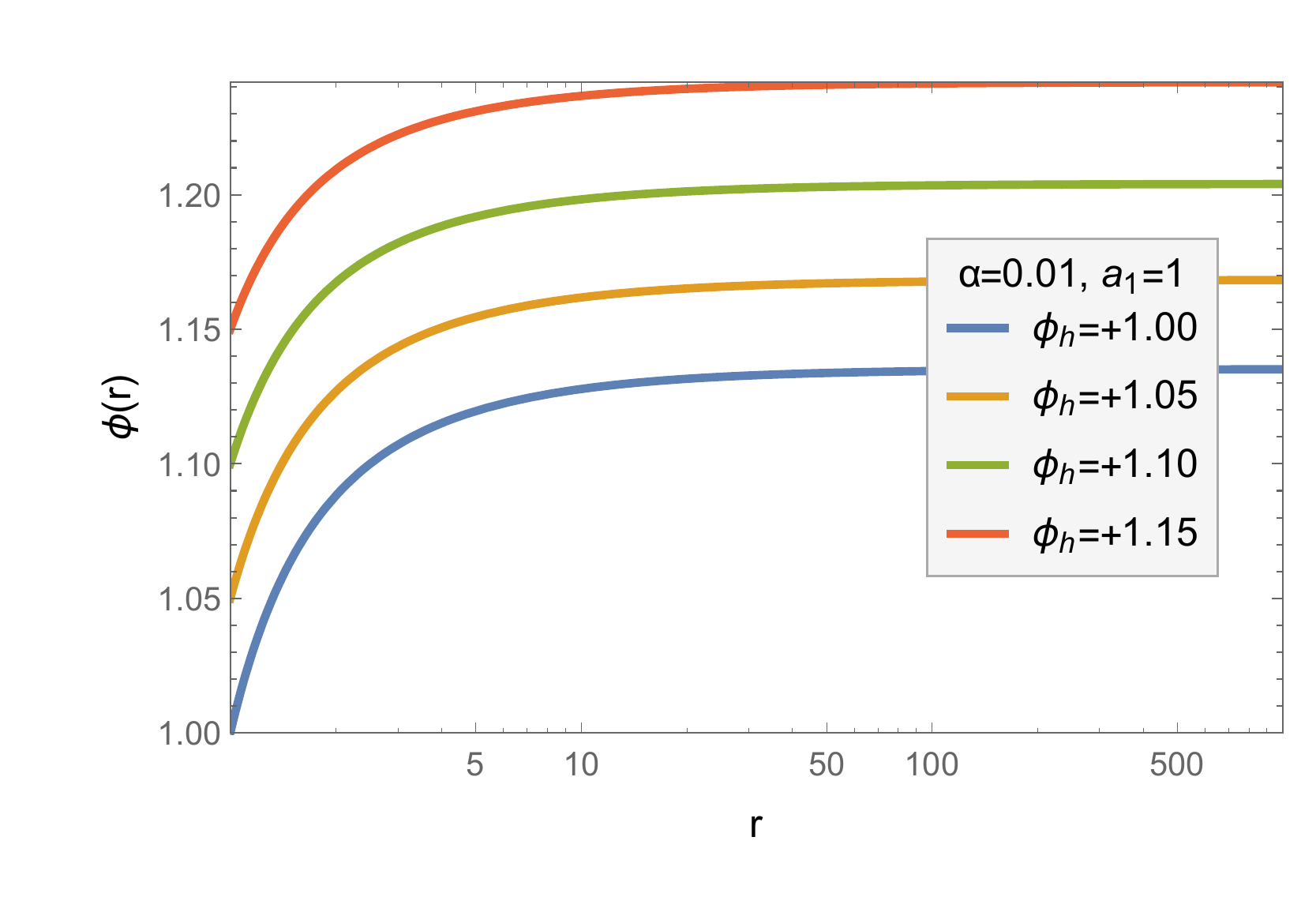}
\endminipage\hfill \hspace*{-1.5cm}
\minipage{0.52\textwidth}
  \includegraphics[width=\linewidth]{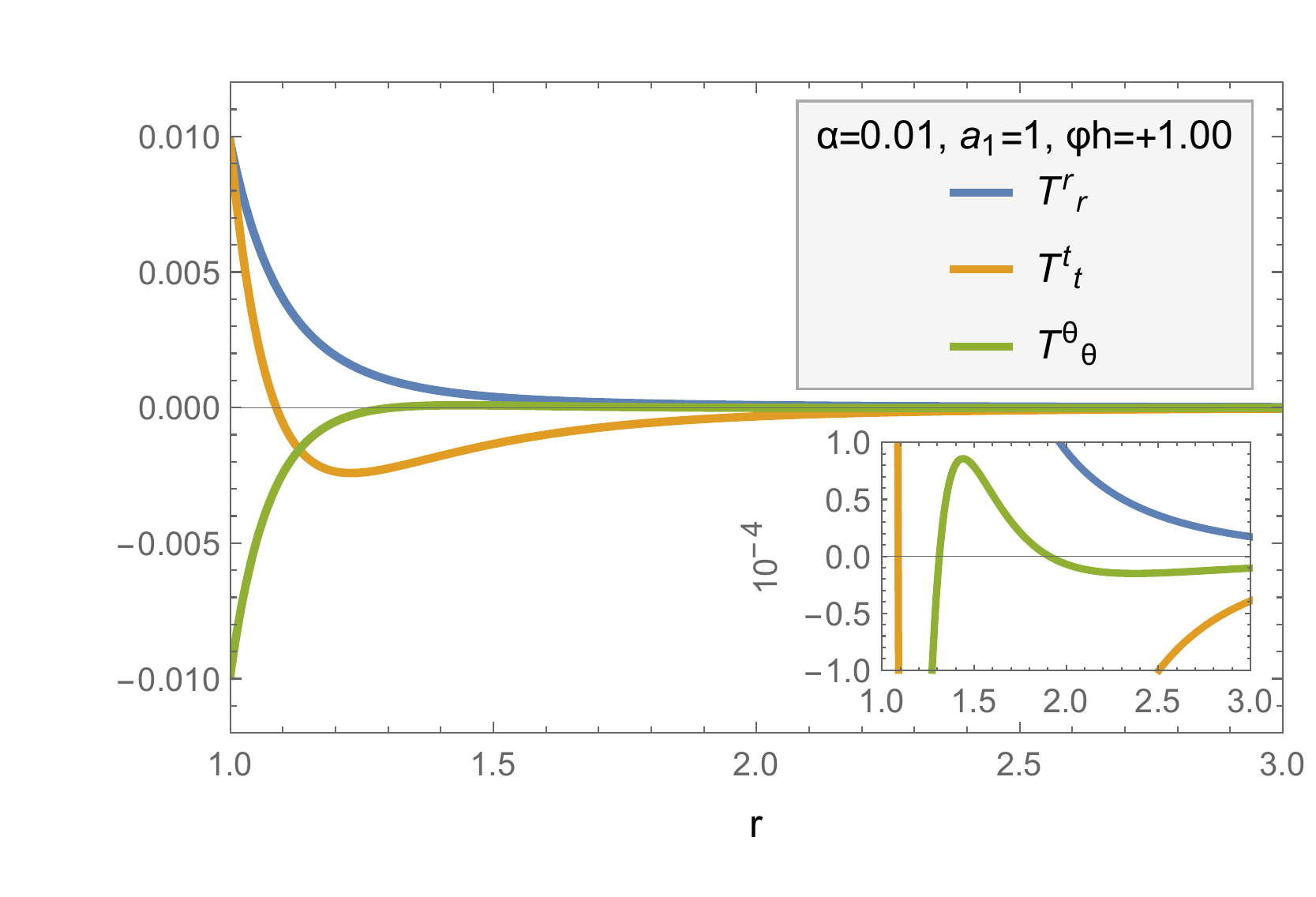}
\endminipage\hfill
\caption{The scalar field $\phi$ (left plot), and the energy-momentum tensor
$T_{\mu\nu}$ (right plot) in terms of the radial coordinate $r$, for $f(\phi)=\alpha/\phi^2$.}
\label{Phi_neg2}
\end{figure}

\begin{itemize}
\item $k=1$: In this case, the constraint for the evasion of the novel no-hair
theorem becomes: $\dot f \phi'=-2\alpha \phi'/\phi^2<0$ which demands $\phi'_h>0$
for all solutions. At the left plot of
Fig. \ref{Phi_neg1}, we present a family of solutions for the scalar field emerging
for this coupling function. All solutions are increasing away from the black-hole
horizon in accordance to the above comment. The components of the energy-momentum
tensor are also well behaved, as may be seen from the right plot of Fig. \ref{Phi_neg1}.
As in the case of the odd polynomial function, an additional set of solutions arises
with $\phi_h<0$ with similar characteristics. 

\item $k=2$: In this case, the constraint becomes: $\dot f \phi'=-\alpha \phi'/\phi^3<0$
which demands $\phi_h \phi'_h>0$. A family of solutions for the scalar field emerging
for this coupling function, with $\phi_h>0$ and increasing with $r$, are presented at
the left plot of Fig. \ref{Phi_neg2} -- a complementary family of solutions with $\phi_h<0$
and decreasing away from the black-hole horizon were also found. The components of the
energy-momentum tensor for an indicative solution are depicted at the right plot of
Fig. \ref{Phi_neg2}, and clearly remain finite over the whole exterior regime.

\end{itemize}

\begin{figure}[b!]
\minipage{0.50\textwidth}
  \includegraphics[width=\linewidth]{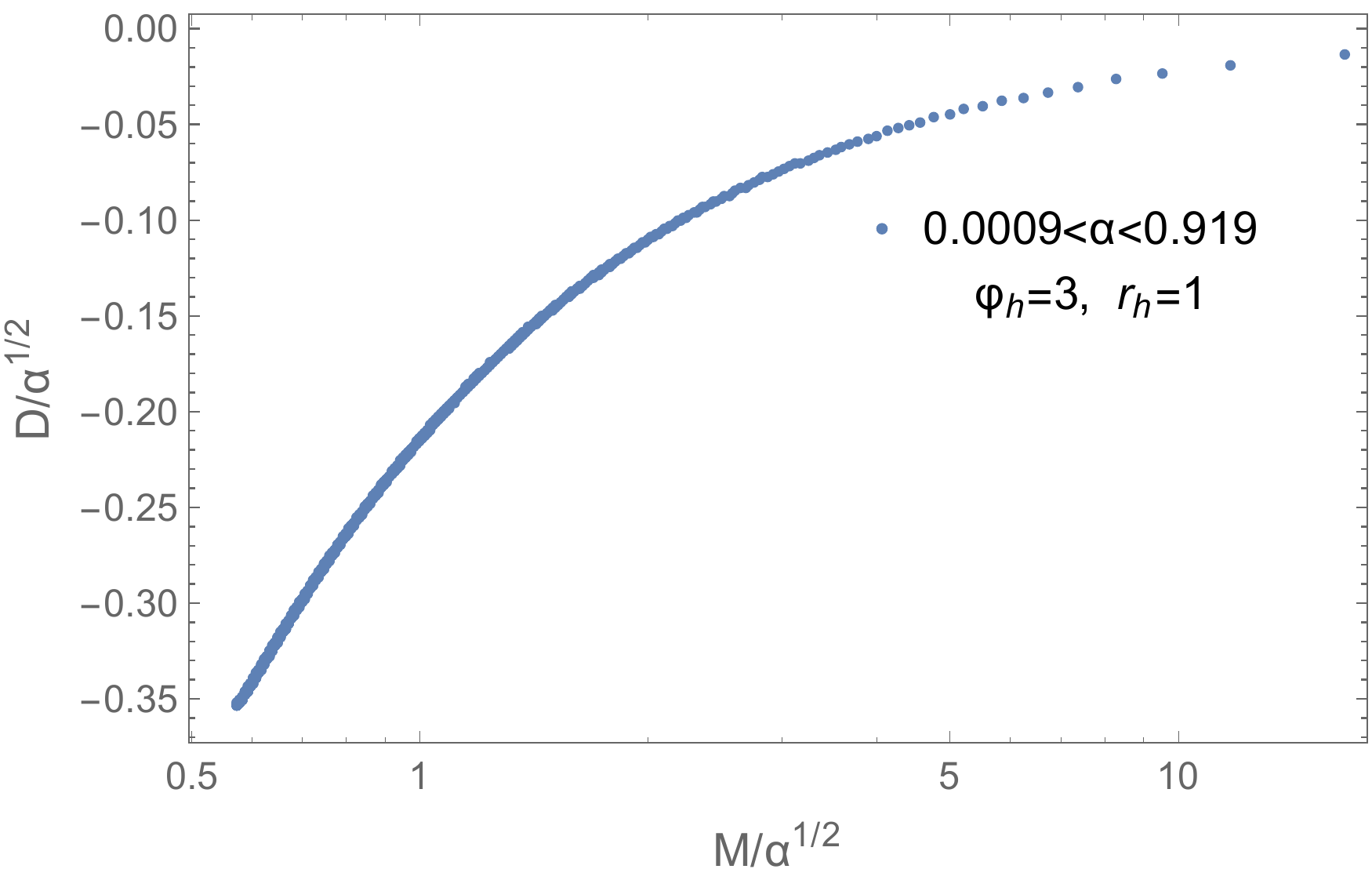}
\endminipage\hfill \hspace*{-2.0cm}
\minipage{0.51\textwidth}
  \includegraphics[width=\linewidth]{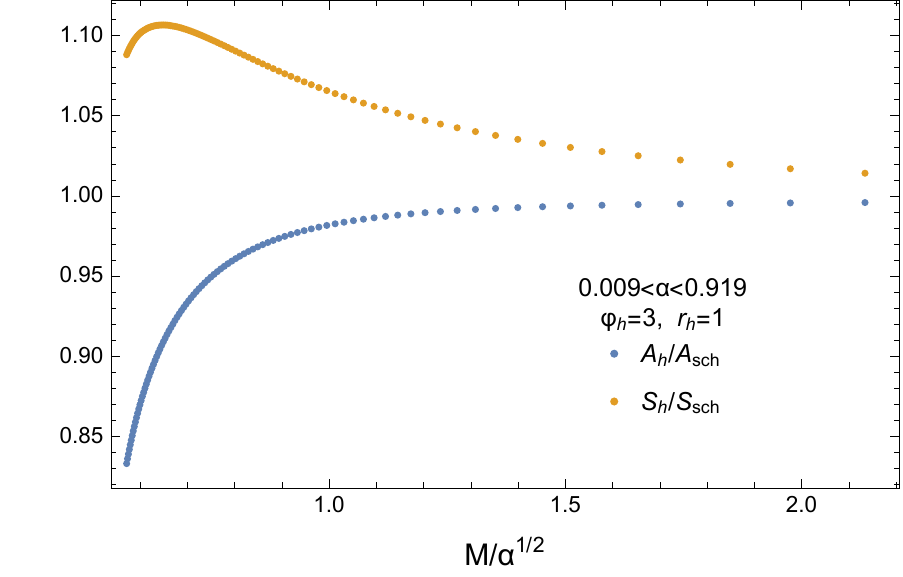}
\endminipage\hfill
\caption{The scalar charge $D$ (left plot), and the ratios $A_h/A_{Sch}$ and $S_h/S_{Sch}$ 
(right plot) in terms of the mass  $M$, for $f(\phi)=\alpha/\phi$.}
\label{D-AS-inv_lin}
\end{figure}

Concerning the characteristics of the `inverse polynomial' GB black holes, we find
again an interesting behaviour -- in Fig. \ref{D-AS-inv_lin}, we depict the inverse
linear case, as an indicative one. The scalar charge $D$ exhibits a monotonic
decreasing behaviour in terms of the mass $M$, as one may see in the left plot
of the figure, approaching zero at the large-mass limit.
In terms of the input parameter $\phi_h$, the scalar charge presents the
anticipated behaviour (and thus is not shown here): for an inverse coupling function,
$D$ increases as the value of $\phi_h$, and thus of the GB coupling, decreases. 
The quantities $A_h$ and $S_h$ increase once again quickly with the mass $M$,
as in Fig. \ref{AS-exp}. The ratio $A_h/A_{Sch}$ is shown in the right plot of
Fig. \ref{D-AS-inv_lin}, and reveals again the constantly smaller size of the GB
black holes compared to the asymptotic Schwarzschild solution, as well as the
existence of a lowest-mass solution. The ratio $S_h/S_{Sch}$, depicted also
in Fig. \ref{D-AS-inv_lin}, reveals that the entire class of these black-hole
solutions  -- independently of their mass -- have a higher entropy compared to
the asymptotic Schwarzschild solution. This feature is in fact unique for the
inverse linear function; a similar analysis for the inverse quadratic coupling
function has produced similar results for the quantities $D(M)$, $D(\phi_h)$,
$A_h/A_{Sch}$ and $S_h/S_{Sch}$ with the only difference being that the very-low
mass regime of the `inverse quadratic' GB black holes have a lower entropy than
the Schwarzschild solution, i.e. the situation resembles more the one depicted in
Fig. \ref{AS-exp}.


\subsection{Logarithmic Function}

\begin{figure}[b!]
\minipage{0.52\textwidth}
  \includegraphics[width=\linewidth]{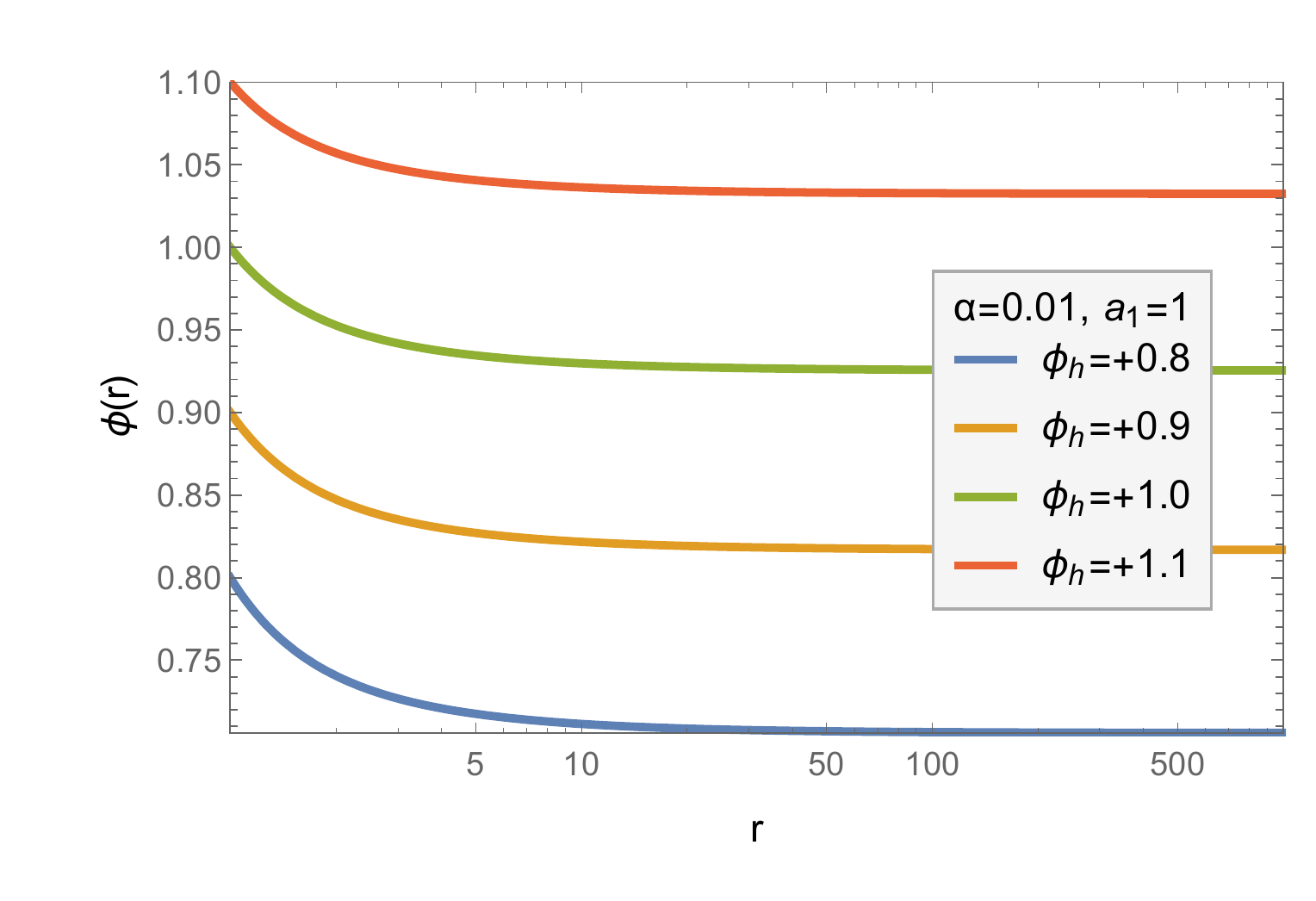}
\endminipage\hfill \hspace*{-1.8cm}
\minipage{0.52\textwidth}
  \includegraphics[width=\linewidth]{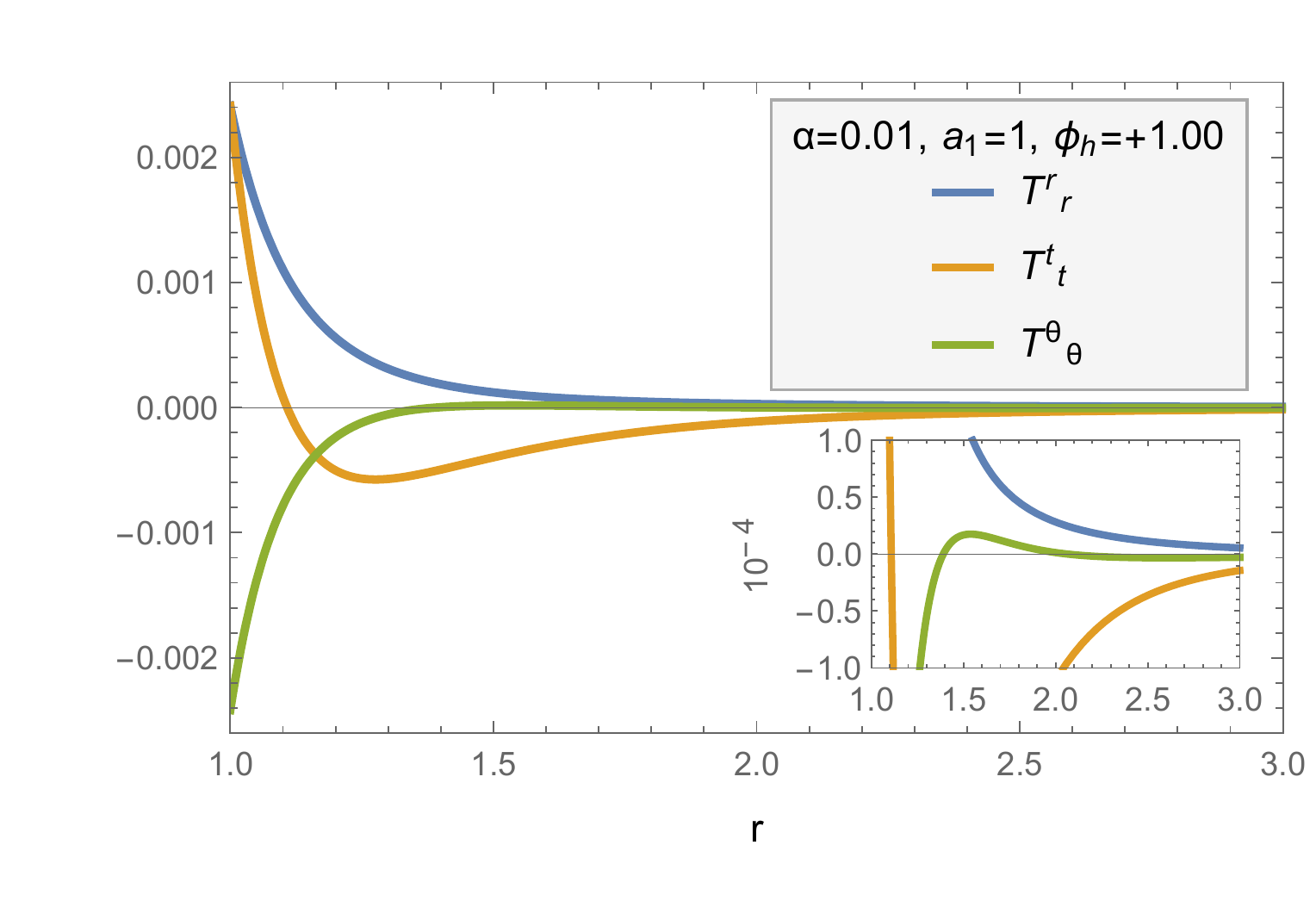}
\endminipage\hfill
\caption{The scalar field $\phi$ (left plot), and the energy-momentum tensor
$T_{\mu\nu}$ (right plot) in terms of the radial coordinate $r$, for $f(\phi)=\alpha\,\ln\phi$.}
\label{Phi_log}
\end{figure}


We finally address the case where $f(\phi)=\alpha \ln(\phi)$. Here, black-hole solutions
emerge for $\dot f \phi' =\alpha \phi'/\phi<0$, near the black-hole horizon; for $\alpha>0$,
this translates to $\phi'_h<0$ (since the argument of the logarithm must be a positive number,
i.e. $\phi>0$). As a result, the solutions for the scalar field are restricted to have a decreasing
behaviour as we move away from the black-hole horizon - this is indeed the behaviour
observed in the class of solutions depicted at the left plot of Fig. \ref{Phi_log}. One may
also observe that the plot includes solutions with either $\phi_h>1$ or $\phi_h<1$, 
or equivalently with $f>0$ or $f<0$. Once again, the old no-hair theorem is proven to
be inadequate to exclude the presence of regular black holes with scalar hair even
in subclasses of the theory (\ref{action}). In contrast, the derived solutions continue to
satisfy the constraints for the evasion of the novel no-hair theorem.  

The components of the energy-momentum tensor are presented in the right plot
of Fig. \ref{Phi_log}: they exhibit the same characteristics as in the cases
presented in the previous subsections with the most important being the monotonic,
decreasing profile of the $T^r_{\,\,r}$ component. The coupling constant $\alpha$
can also take a variety of values as long as it satisfies Eq. (\ref{con-f}); in
this case, the monotonic behaviour of $\phi$ over the whole exterior space of
the black hole is preserved independently of the value of $\alpha$. The
energy-momentum tensor also assumes the same form as in Fig. \ref{Phi_log},
and thus we refrain from presenting any new plots.

\begin{figure}[t!]
\minipage{0.50\textwidth}
  \includegraphics[width=\linewidth]{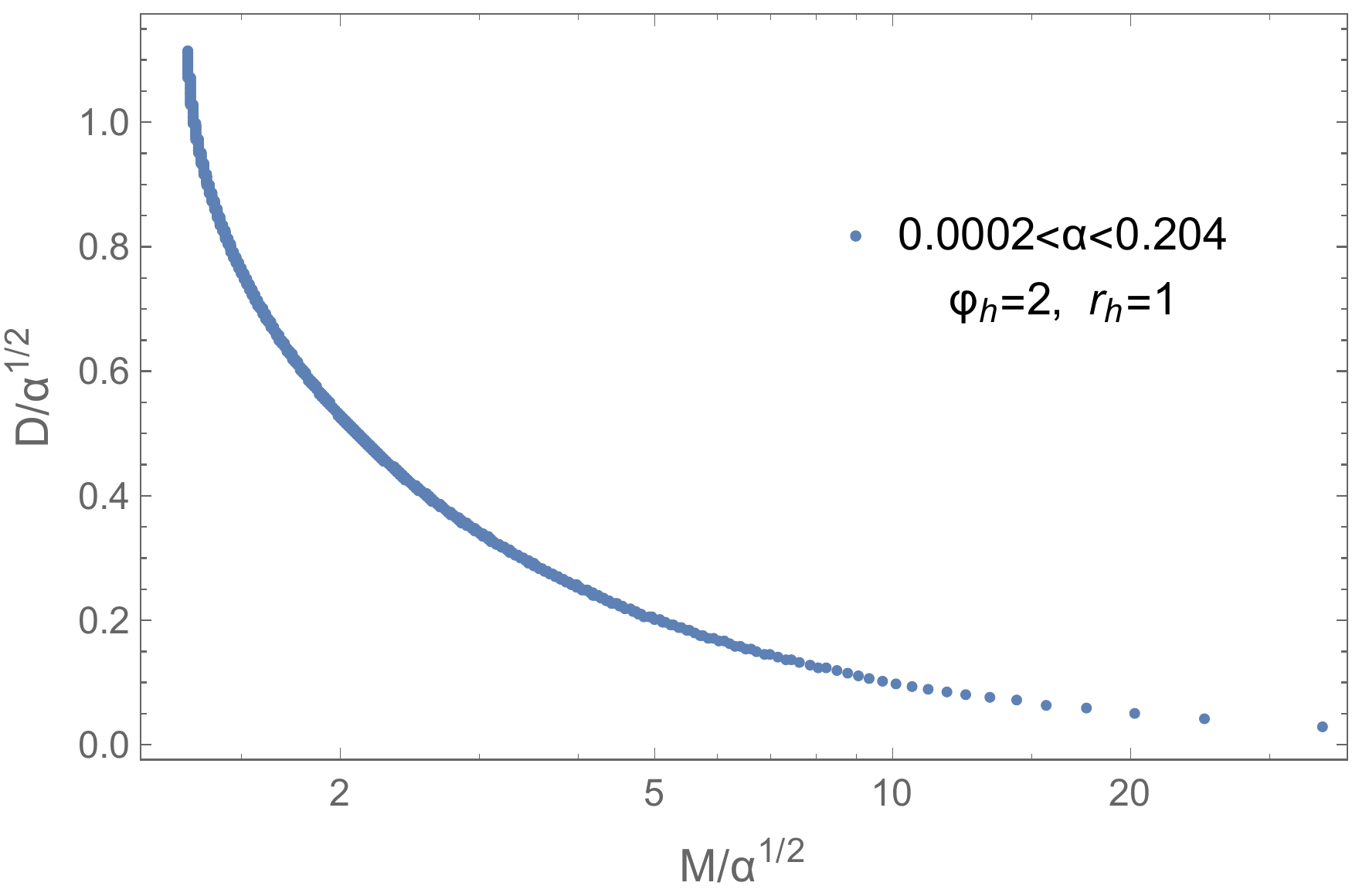}
\endminipage\hfill \hspace*{-2.0cm}
\minipage{0.51\textwidth}
  \includegraphics[width=\linewidth]{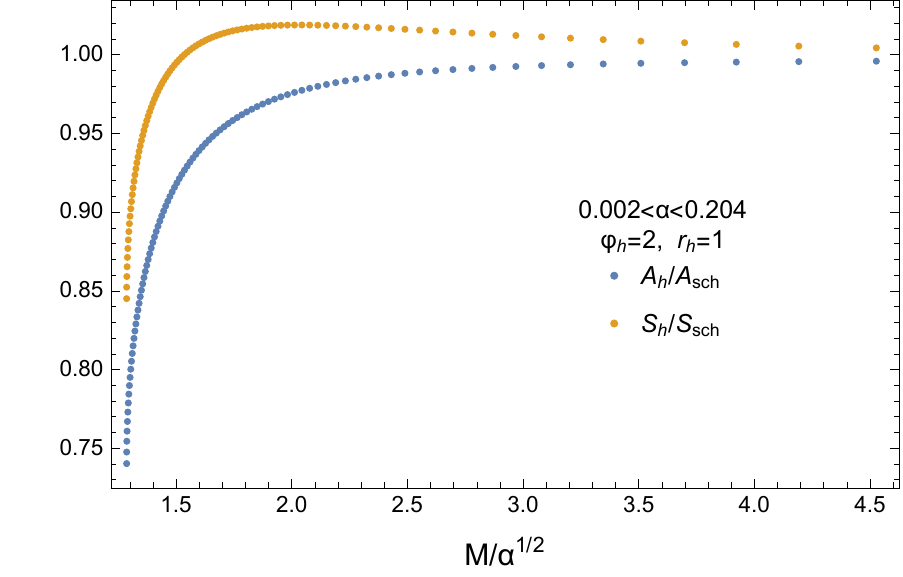}
\endminipage\hfill
\caption{The scalar charge $D$ (left plot), and the ratios $A_h/A_{Sch}$ and $S_h/S_{Sch}$ 
(right plot) in terms of the mass  $M$, for $f(\phi)=\alpha\,\ln\phi$.}
\label{D-AS-log}
\end{figure}

The scalar charge $D$ of the `logarithmic' GB black holes, in terms of the mass
$M$ is shown in the left plot of Fig. \ref{D-AS-log}. Once again, we observe that,
as the mass of the black-hole solution increases, $D$ decreases towards a
vanishing  value -- our previous analysis has shown that this feature
is often connected with the thermodynamical stability of the solutions. Indeed,
as it may be seen from
the right plot of Fig. \ref{D-AS-log}, the ratio $S_h/S_{Sch}$ is above unity for
a very large part of the mass-regime; it is again the small-mass regime that
is excluded from the thermodynamical stable solutions. The scalar charge $D$
has again the anticipated behaviour in terms of the parameter $\phi_h$: since
it is the $f'(\phi)=\alpha/\phi$ that enters the scalar equation (\ref{phi-eq_0}),
the scalar charge increases as $\phi_h$ decreases. The ratio $A_h/A_{Sch}$ is
also shown in the right plot of Fig. \ref{D-AS-log}: as in the previous cases,
it corresponds to a class of smaller black-hole solutions, compared to the
Schwarzschild solution of the same mass, with a minimum-allowed mass that
approaches the asymptotic Schwarzschild solution as $M$ reaches large values.


\section{Conclusions}

The emergence of regular black-hole solutions in theories with scalar fields
has always attracted the interest of researchers especially in conjunction with
the no-hair theorems. The latter were formed in an attempt to limit the emergence
of black-hole solutions in the context of theories beyond General Relativity.
However, a small number of theories, in which the evasion of those theorems was
realised, were found in the literature over the years leading to novel solutions
with scalar hair. 

A main characteristic of those theories was the coupling of the scalar field
with higher-order gravitational terms. When one focuses on theories where the
higher-curvature term is the quadratic Gauss-Bonnet term, the number of
black-hole solutions with non-trivial scalar hair, that have been found in
the literature, is very limited \cite{DBH, Benkel}. These two works
considered the particular cases of a string-inspired, exponential coupling
function between the scalar field and the GB term, and a shift-symmetric
linear coupling function, respectively. In a previous work of ours \cite{ABK1},
we have nevertheless demonstrated that regular black-hole solutions with scalar
hair may be constructed in the context of a much more general class of
theories that contain the GB term. For this to realise and the theory
to evade the no-hair theorems, the general coupling function $f(\phi)$ had
to satisfy two constraints. The first constraint is imposed on the value
of $\phi'_h$; this uniquely determines this quantity which may then be used
as an input parameter for the construction of the solutions. Using this,
we were indeed able to produce asymptotically-flat, regular black-hole
solutions; these were briefly presented in \cite{ABK1}.

In the present work, we have extended the analysis of \cite{ABK1} by considering
several subclasses of the general theory that contained the Ricci scalar, a scalar
field and the GB term. We have studied a large number of choices for the
coupling function $f(\phi)$ between the scalar field and the GB term:
exponential, polynomial (even and odd), inverse polynomial (even and odd)
and logarithmic. In each case, employing the constraint (\ref{con-phi'})
for the value of $\phi'_h$, we constructed a large number of exact black-hole
solutions with scalar hair, and studied in detail their characteristics. 
Our solutions
were characterised by a universal behaviour of the components of the metric
tensor, having the expected behaviour near the black-hole horizon and
asymptotic flatness at radial infinity. All curvature invariant quantities
were examined, and found to have a similar universal profile, independently
of the form of the coupling function $f$, that ensured the finiteness of 
the space-time and thus the regularity of all solutions. 

The same regularity characterised the components of the energy-momentum
tensor over the whole radial regime. In fact, the first constraint on
the value of $\phi'_h$ -- necessary for the evasion of the novel no-hair
theorem -- simultaneously guarantees the regularity of the scalar field
at the black-hole horizon and therefore, the regularity of the solution.
The second constraint for the evasion of the no-hair theorem involves
both $\phi'_h$ and $\phi''_h$, and determines the behaviour of 
$T^r_{\,\,r}$ near the black-hole horizon; this constraint was automatically
satisfied by all the constructed solutions and demanded no further action
or fine-tuning of the parameters. It is worth noting that these constraints
were local as they applied to the black-hole radial regime, and were
therefore easy to check. On the other hand, the old
no-hair theorem, based on an integral constraint over the whole radial
regime, fails to lead to a unique constraint whose violation or not
would govern the existence of regular black-hole solutions. A special
form of it, i.e. $f(\phi)>0$, was found to lead indeed to novel solutions
when satisfied but it could not exclude their emergence in the opposite case.

The profile for the scalar field $\phi$ was found in all of the cases considered,
and was indeed regular over the entire radial domain. The scalar field
increased or decreased away from the black-hole horizon, always in accordance
with the constraint (\ref{con-phi'}), and approached a constant value at
asymptotic infinity. The scalar charge $D$ was determined in each case, and
its dependence on either $\phi_h$ or $M$ was studied. In terms of the first
parameter, whose value determines the magnitude of the GB coupling in the
theory, its behaviour was the anticipated one: for large values of that coupling,
$D$ assumed a large value, while for small couplings $D$ tended to zero. 
Its profile in terms of the mass $M$ of the black hole had a universal
behaviour in the large-$M$ limit, with $D$ assuming increasingly smaller values,
however, in the small-$M$ limit, each family of solutions presented a different
behaviour (either monotonic or not monotonic). In all cases, however, the
scalar charge is an $M$-dependent quantity, and therefore our solutions have
a non-trivial scalar field but with a ``secondary'' hair.

The horizon area $A_h$ and entropy $S_{h}$ of the solutions were also found
in each case. Both quantities quickly increase with the mass $M$, with the
former always dominating over the latter. The function $A(M)$ also revealed
a generic feature of all black-hole solutions found, namely the existence
of a lower value for the horizon radius $r_h$ and thus of the horizon
area $A_h$ of the black hole; this is due to the constraint (\ref{con-f})
that, for fixed $\phi_h$ and parameter $\alpha$, does not allow for regular
solutions with horizon radius smaller than a minimum value, given by 
$(r_h^2)_{min} = 4\sqrt{6}\,|\dot f_h|$, to emerge. This, in turn, imposes
a lower-bound on the mass of the black hole solution, and therefore all
curves $A(M$) terminate at a specific point in the low-mass regime. 

The study of the ratios $A_h/A_{Sch}$ and $S_h/S_{Sch}$, with respect to the
corresponding quantities of the Schwarzschild solution with the same mass,
had even more information to offer. The first ratio remained below unity
for all classes of black-hole solutions found and for all mass regimes;
as a result, we may conclude that the presence of the additional, gravitational
GB term leads to the formation of more compact black holes compared to the
standard General Relativity (GR). In the large-mass limit, the horizon area
of all black-hole solutions approached the Schwarzschild value -- the same
was true for the entropy ratio $S_h/S_{Sch}$; these two features together
suggest that, for large masses, it will be extremely difficult to distinguish
between GB black holes and their GR analogues.

Do we really expect to detect any of these classes of GB black holes in the
universe? This depends firstly on their stability behaviour, a topic that
needs to be studied carefully and individually for each class of solutions
presented in this work. The curves $S_h/S_{Sch}(M)$, that we produced, may
provide hints for their stability: as mentioned above, the entropy of all
solutions found here approached, in the large-$M$ limit, the Schwarzschild value
thus it is quite likely that large GB black holes share the stability of 
the Schwarzchild solution. For smaller masses, where the GB black
holes are expected to differ from their GR analogue, different profiles
were observed: the `exponential', `inverse-quadratic' and `logarithmic' 
GB black holes had a ratio $S_h/S_{Sch}$ larger than unity for the entire
intermediate mass-regime but smaller than unity in the very-low-mass regime.
These results point towards the thermodynamical stability of solutions
with intermediate and large masses but to an instability for solutions
with small masses (although even in the latter case, an accretion of mass
from their environment could lead to an increase in their mass and to a
change in their (in)stability). On the other hand, the `quadratic', `quartic'
and `linear' GB black holes had their entropy ratio $S_h/S_{Sch}$ below
unity over the entire mass-regime, and perhaps do not lead to stable
configurations. Finally, two classes of solutions, the `inverse-linear'
and the first branch of the `cubic' GB black holes have their ratio
$S_h/S_{Sch}$ larger than unity for all values of the black-hole mass -
small, intermediate and large - and may hopefully lead to stable solutions
with a variety of masses. In all cases, a careful study of all the above
solutions under perturbations is necessary in order to verify or refute
the above expectations (the only class of GB black-hole solutions that 
have been studied under linear perturbations are the `exponential' ones
that were found to be indeed stable \cite{DBH} in accordance to the
above comments).

Assuming therefore that one or more classes of the aforementioned black-hole
solutions are stable, we then need a number of signatures or observable
effects that would distinguish them from their GR analogues and convince
us of their existence. A generic feature of all GB black holes is their
minimum horizon radius: if, in the small-mass limit, certain families of
GB black holes are more favourable to emerge compared to the Schwarzschild
solution -- from the stability point-of-view, then the observed black holes
will not have an arbitrarily small mass. Also, in the small-mass limit,
observable effects may include deviations from standard GR in the calculation
of the bending angle of light, the precession observed in near-horizon
orbits and the spectrum from their accretion discs. Studies of this type
have been performed \cite{Chakra} for black holes in the
Einstein-scalar-GB theory with a linear coupling \cite{SZ} -- a special
case of our analysis -- and shown that the near-horizon strong dynamics
may leave its imprint on all of these observables.

Our GB black-hole solutions are characterized also by a scalar charge. A
previous analysis of dilatonic (`exponential') GB black holes \cite{Kunz} has
revealed that scalar radiation is rather suppressed, especially for
non-spinning black holes, unless particular couplings are introduced in
the theory between the scalar field and ordinary matter. In addition, in
\cite{Yagi} it was demonstrated that the scalar charge of neutron stars,
emerging in the context of the same theory, is extremely small. We would
like to add to this that, according to our analysis, the more stable
configurations tend to correspond to black holes with small scalar charge.
Perhaps, future observations of gravitational waves from black-hole 
or neutron-star processes could lead to clear signatures (or impose
constraints) on the existence of GB compact objects provided that these
objects have a small mass and/or a large scalar charge. Finally, the
measurement of the characteristic frequencies of the quasi-normal modes
(especially the polar sector) will also help to distinguish these solutions
from their GR analogues \cite{Kunz}. 


{\bf Note added:} While this manuscript was at the last stages of preparation, two
additional works appeared \cite{Doneva, Silva} where the emergence of black-hole
solutions with scalar hair in the context of the Einstein-scalar-Gauss-Bonnet theory
was studied.

{\bf Acknowledgments}
G.A. would like to thank Onassis Foundation for the financial
support provided through its scholarship program.

\appendix

\numberwithin{equation}{section}

\section{Set of Differential Equations}

Here, we display the explicit expressions of the coefficients $P$, $Q$ and $S$ that
appear in the system of differential equations (\ref{A})-(\ref{phi}) whose solution determines
the metric function $A$ and the scalar field $\phi$. They are:
\bea
P&=&+e^{4B}\Bigl(32A'\dot{f}-48r A'^2\dot{f}-8r^2\phi'-4r^3A'\phi'-64r\phi'^2\dot{f}\Bigr)
+e^{3B}\Bigl(-64A'\dot{f}+96rA'^2\dot{f}\nonumber\\
&&+48r^2A'^3\dot{f}+8r^2\phi'-4r^3A'\phi'-4r^4A'^2\phi'+128A'^2\phi'\dot{f}^2
+96rA'^3\phi'\dot{f}^2+64r\phi'^2\dot{f}\nonumber\\
&&+24r^2A'\phi'^2\dot{f}-20r^3A'^2\phi'^2\dot{f}-2r^4\phi'^3+96rA'\phi'^3\dot{f}^2
-16r^3\phi'^4\dot{f}+32r^2\phi'^3\ddot{f}\nonumber\\
&&+r^5A'\phi'^3 +16r^3A'\phi'^3\ddot{f}\Bigr)
+16e^{2B}\Bigl(8A'\dot{f}-12rA'^2\dot{f}-20r^2A'^3\dot{f}-64A'^2\phi'\dot{f}^2\nonumber\\
&&-112rA'^3\phi'\dot{f}^2 -14r^2A'\phi'^2\dot{f}+19r^3A'^2\phi'^2\dot{f}-
96A'^3\phi'^2\dot{f}^3-32rA'\phi'^3\dot{f}^2\nonumber\\
&&+36r^2A'^2\phi'^3\dot{f}^2 +8r^3\phi'^4\dot{f}-4r^4A'\phi'^4\dot{f}
-8r^2\phi'^3\ddot{f}+4r^3A'\phi'^3\ddot{f}-32r^2A'\phi'^4\dot{f}\ddot{f}\Bigr)\nonumber\\
&&+16e^B\Bigl(8A'^2\phi'\dot{f}^2+38rA'^3\phi'\dot{f}^2+64A'^3\phi'^2\dot{f}^3+
18rA'\phi'^3\dot{f}^2-17r^2A'^2\phi'^3\dot{f}^2\Bigr)\nonumber\\
&&-1152A'^3\phi'^2\dot{f}^3\,,\label{P}
\eea
\bea
Q&=&+32e^{5B}r-e^{4B}\left(64r+24r^2A'+160\phi'\dot{f}+48rA'\phi'\dot{f}+4r^3\phi'^2+128r\phi'^2\ddot{f}\right)\nonumber\\
&&+e^{3B}\left(32r+24r^2A'-8r^3A'^2+320\phi'\dot{f}+224rA'\phi'\dot{f}-32r^2A'^2\phi'\dot{f}-
12r^3\phi'^2\right.\nonumber\\
&&\left.+6r^4A'\phi'^2+256A'\phi'^2\dot{f}^2-32rA'^2\phi'^2\dot{f}^2-24r^2\phi'^3\dot{f}+12r^3A'\phi'^3\dot{f}
-32r\phi'^4\dot{f}^2\right.\nonumber\\
&&\left.-r^5\phi'^4+256r\phi'^2\ddot{f}+128r^2A'\phi'^2\ddot{f}+640\phi'^3\dot{f}\ddot{f}+256rA'\phi'^3\dot{f}\ddot{f}-16r^3\phi'^4\ddot{f}\right)\nonumber\\
&&+e^{2B}\left(128r^2A'^2\phi'\dot{f}-160\phi'\dot{f}
-176rA'\phi'\dot{f}-640A'\phi'^2\dot{f}^2+320rA'^2\phi'^2\dot{f}^2\right.\nonumber\\
&&\left.+152r^2\phi'^3\dot{f}-52r^3A'\phi'^3\dot{f}+128A'^2\phi'^3\dot{f}^3+256\phi'^4r\dot{f}^2-80r^2A'\phi'^4\dot{f}^2+4r^4\phi'^5\dot{f}\right.\nonumber\\
&&\left.-128r\phi'^2\ddot{f}-128r^2A'\phi'^2\ddot{f}-1280\phi'^3\dot{f}\ddot{f}-1280rA'\phi'^3\dot{f}\ddot{f}+16r^3\phi'^4\ddot{f}\right.\nonumber\\
&&\left.-1280A'\phi'^4\dot{f}^2\ddot{f}  +64r^2\phi'^5\dot{f}\ddot{f}\right)+e^B\left(384A'\phi'^2\dot{f}^2-672rA'^2\phi'^2\dot{f}^2-768A'^2\phi'^3\dot{f}^3\right.\nonumber\\
&&\left.-480r\phi'^4\dot{f}^2+144r^2A'\phi'^4\dot{f}^2+640\phi'^3\dot{f}\ddot{f}+
1024rA'\phi'^3\dot{f}\ddot{f}+3584A'\phi'^4\dot{f}^2\ddot{f}\right.\nonumber\\
&&\left.-64r^2\phi'^5\dot{f}\ddot{f}\right)+1152A'^2\phi'^3\dot{f}^3-
2304A'\phi'^4\dot{f}^2\ddot{f}\,,\label{Q}\nonumber
\eea
\bea
S &=& +128r\dot{f}e^{4B}+8e^{3B}\left(r^4\phi'-32r\dot{f}-16r^2A'\dot{f}-80\phi'\dot{f}^2
32rA'\phi'\dot{f}^2+4r^3\phi'^2\dot{f}\right)\nonumber\\
&&+32e^{2B}\left(4r\dot{f}+4r^2A'\dot{f}+40\phi'\dot{f}^2+40rA'\phi'\dot{f}^2-
3r^3\phi'^2\dot{f}+40A'\phi'^2\dot{f}^3-4r^2\phi'^3\dot{f}^2\right)\nonumber\\
&&+8e^B\left(32r^2\phi'^3\dot{f}^2-80\phi'\dot{f}^2-
128rA'\phi'\dot{f}^2-448A'\phi'^2\dot{f}^3\right)+2304A'\phi'^2\dot{f}^3\,.\label{S}
\eea

\section{Scalar Quantities}

By employing the metric components of the line-element (\ref{metric}), one may
compute the following scalar-invariant gravitational quantities:
\bea
R&=&+\frac{e^{-B}}{2r^2}\left(4e^B-4-r^2A'^2+4rB'-4rA'+r^2A'B'-2r^2A''\right),\label{A1}\\\nonumber\\
R_{\mu\nu}R^{\mu\nu}&=&+\frac{e^{-2B}}{16 r^4}\left[8(2-2e^B+rA'-rB')^2+r^2(rA'^2-4B'-rA'B'+2rA'')^2\right.\nonumber\\
&&\left.+r^2(rA'^2+A'(4-rB')+2rA'')^2\right],\\\nonumber\\
R_{\mu\nu\rho\sigma}R^{\mu\nu\rho\sigma}&=&+\frac{e^{-2B}}{4r^4}\left[r^4A'^4-2r^4A'^3B'-4r^4A'B'A''+r^2A'^2(8+r^2B'^2+4r^2A'')\right.\nonumber\\
&&\left.+16(e^B-1)^2+8r^2B'^2+4r^4A''^2\right],\\\nonumber\\
R_{GB}^2&=&+\frac{2e^{-2B}}{r^2}\left[(e^B-3)A'B'-(e^B-1)A'^2-2(e^B-1)A''\right].\label{A4}
   \eea


\end{document}